\documentclass[twocolumn]{aastex631}
\usepackage{amsmath}

\shorttitle{Hard X-ray Flares in NICER Monitoring of AT2019cuk}
\shortauthors{Masterson et al.}

\begin{document}

\title{Unusual Hard X-ray Flares Caught in NICER Monitoring of the Binary Supermassive Black Hole Candidate AT2019cuk/Tick~Tock/SDSS~J1430+2303}

\author[0000-0003-4127-0739]{Megan Masterson}
\affiliation{MIT Kavli Institute for Astrophysics and Space Research,
Massachusetts Institute of Technology, 
Cambridge, MA 02139, USA}
	
\author[0000-0003-0172-0854]{Erin Kara}
\affiliation{MIT Kavli Institute for Astrophysics and Space Research,
Massachusetts Institute of Technology, 
Cambridge, MA 02139, USA}
	
\author[0000-0003-1386-7861]{Dheeraj R. Pasham}
\affiliation{MIT Kavli Institute for Astrophysics and Space Research,
Massachusetts Institute of Technology, 
Cambridge, MA 02139, USA}

\author[0000-0002-1271-6247]{Daniel J. D’Orazio}
\affiliation{Niels Bohr International Academy, Niels Bohr Institute, Blegdamsvej 17, DK-2100 Copenhagen, Denmark}

\author[0000-0001-5819-3552]{Dominic J. Walton}
\affiliation{Institute of Astronomy, 
University of Cambridge, 
Madingley Road, Cambridge CB3 0HA, UK}
\affiliation{Centre for Astrophysics Research, University of Hertfordshire, College Lane, Hatfield AL10 9AB, UK}

\author[0000-0002-9378-4072]{Andrew C. Fabian}
\affiliation{Institute of Astronomy, 
University of Cambridge, 
Madingley Road, Cambridge CB3 0HA, UK}

\author[0000-0002-2235-3347]{Matteo Lucchini}
\affiliation{MIT Kavli Institute for Astrophysics and Space Research,
Massachusetts Institute of Technology, 
Cambridge, MA 02139, USA}

\author[0000-0003-4815-0481]{Ronald A. Remillard}
\affiliation{MIT Kavli Institute for Astrophysics and Space Research,
Massachusetts Institute of Technology, 
Cambridge, MA 02139, USA}

\author{Zaven Arzoumanian}
\affiliation{X-Ray Astrophysics Laboratory, NASA Goddard Space Flight Center, Code 662, Greenbelt, Maryland 20771, USA}

\author[0000-0003-1169-6763 ]{Otabek Burkhonov}
\affiliation{Ulugh Beg Astronomical Institute, Astronomy Street 33, Tashkent 100052, Uzbekistan}

\author[0000-0003-4422-6426]{Hyeonho Choi}
\affiliation{SNU Astronomy Research Center, Astronomy Program, Department of Physics and Astronomy, Seoul National University, 1 Gwanak-ro, Gwanak-gu, Seoul, 08826, Republic of Korea}

\author[0000-0001-9730-3769]{Shuhrat A. Ehgamberdiev}
\affiliation{Ulugh Beg Astronomical Institute, Astronomy Street 33, Tashkent 100052, Uzbekistan}
\affiliation{National University of Uzbekistan, Tashkent 100174, Uzbekistan}

\author[0000-0001-7828-7708]{Elizabeth C. Ferrara}
\affil{Department of Astronomy, University of Maryland, College Park, MD, 20742, USA}
\affil{Center for Exploration and Space Studies (CRESST), NASA/GSFC, Greenbelt, MD 20771, USA}
\affil{NASA Goddard Space Flight Center, Greenbelt, MD 20771, USA}

\author[0000-0002-5063-0751]{Muryel Guolo}
\affiliation{Department of Physics and Astronomy, Johns Hopkins University, 400 N. Charles St., Baltimore, MD 21218, USA}

\author[0000-0002-8537-6714]{Myungshin Im}
\affiliation{SNU Astronomy Research Center, Astronomy Program, Department of Physics and Astronomy, Seoul National University, 1 Gwanak-ro, Gwanak-gu, Seoul, 08826, Republic of Korea}

\author[0000-0002-9532-1653]{Yonggi Kim}
\affiliation{Department of Astronomy and Space Science, Chungbuk National University Observatory, Chungdae-ro 1, Seowon-Gu, Cheongju, Chungbuk 28644, Republic of Korea}

\author[0000-0003-0570-6531]{Davron O. Mirzaqulov}
\affiliation{Ulugh Beg Astronomical Institute, Astronomy Street 33, Tashkent 100052, Uzbekistan}

\author[0000-0002-6639-6533]{Gregory S. H. Paek}
\affiliation{SNU Astronomy Research Center, Astronomy Program, Department of Physics and Astronomy, Seoul National University, 1 Gwanak-ro, Gwanak-gu, Seoul, 08826, Republic of Korea}

\author{Hyun-Il Sung}
\affiliation{Korea Astronomy and Space Science Institute, 776 Daedeokdae-ro, Yuseong-gu, Daejeon 34055, Republic of Korea}

\author{Joh-Na Yoon}
\affiliation{Department of Astronomy and Space Science, Chungbuk National University Observatory, Chungdae-ro 1, Seowon-Gu, Cheongju, Chungbuk 28644, Republic of Korea}

\correspondingauthor{Megan Masterson}
\email{mmasters@mit.edu}

\begin{abstract}
The nuclear transient AT2019cuk/Tick~Tock/SDSS~J1430+2303 has been suggested to harbor a supermassive black hole (SMBH) binary near coalescence. We report results from high-cadence NICER X-ray monitoring with multiple visits per day from January-August 2022, as well as continued optical monitoring during the same time period. We find no evidence of periodic/quasi-periodic modulation in the X-ray, UV, or optical bands, however we do observe exotic hard X-ray variability that is unusual for a typical AGN. The most striking feature of the NICER light curve is repetitive hard (2-4 keV) X-ray flares that result in distinctly harder X-ray spectra compared to the non-flaring data. In its non-flaring state, AT2019cuk looks like a relatively standard AGN, but it presents the first case of day-long, hard X-ray flares in a changing-look AGN. We consider a few different models for the driving mechanism of these hard X-ray flares, including: (1) corona/jet variability driven by increased magnetic activity, (2) variable obscuration, and (3) self-lensing from the potential secondary SMBH. We prefer the variable corona model, as the obscuration model requires rather contrived timescales and the self-lensing model is difficult to reconcile with a lack of clear periodicity in the flares. These findings illustrate how important high-cadence X-ray monitoring is to our understanding of the rapid variability of the X-ray corona and necessitate further high-cadence, multi-wavelength monitoring of changing-look AGN like AT2019cuk to probe the corona-jet connection.
\end{abstract}

\keywords{Active galactic nuclei (16)–High energy astrophysics (739)–Supermassive black holes (1663)–X-ray transient sources (1852)}

\section{Introduction}

In the theory of hierarchical structure growth, supermassive black hole (SMBH) binaries are thought to be a natural result of merging galaxies. Indeed, kiloparsec-separation dual SMBH systems have been identified in merging systems across the electromagnetic spectrum, including in the radio \citep[e.g.][]{Fu2011,Muller-Sanchez2015}, optical \citep[e.g.][]{Liu2018,Silverman2020}, and X-ray \citep[e.g.][]{Komossa2003,Koss2011,Foord2020}. Some close-separation, parsec-scale binaries have also been resolved using very long baseline interferometry \citep[VLBI; e.g.][]{Rodriguez2006,Burke-Spolaor2011}. However, imaging SMBH binaries at sub-parsec separations is difficult due to the extremely high angular resolution required, which is only possible with observatories like the Event Horizon Telescope or its upgrades \citep[e.g.][]{D'Orazio2018b,Kurczynski2022}. Instead, spectroscopic signatures, like double-peaked or systematically shifted emission lines around the accreting black hole \citep[e.g.][]{Tsalmantza2011,Ju2013}, or time-dependent photometric signatures, like periodic modulation of the light curve \citep[e.g.][]{Graham2015a,Graham2015b,Liu2015,Charisi2016,Chen2022}, have been argued to indicate the presence of a close-separation binary SMBH. While spectroscopic searches for binary candidates have been fruitful, this behavior is not unique to binaries, with other explanations including a recoiling SMBH \citep[e.g.][]{Blecha2008,Komossa2008} or Keplerian motion in the disk \citep[e.g.][]{Eracleous1994,Chornock2010,Gaskell2010}. Likewise, finding periodic signals among stochastic AGN variability requires many periods of data (years to decades) to confirm a close-separation binary system \citep[see e.g.][]{Vaughan2005,Vaughan2016}. To date, our best close-separation binary SMBH candidates come from periodic light curves with extremely long temporal baselines \citep[e.g. OJ 287;][]{Sillanpaa1988,Lehto1996,Valtonen2008}.

AT2019cuk (also known as Tick Tock, SDSS~J1430, or ZTF18aarippg) is a nuclear transient discovered by the Zwicky Transient Facility (ZTF) on 2019 March 30 in the host galaxy SDSS J143016.05+230244.4 \citep[$z = 0.081$; ][]{Oh2015}. It rose by approximately one magnitude in the optical $g$-band, reaching a peak magnitude of 17 and then declining with an apparently oscillatory pattern, reminiscent of what one would expect from a binary SMBH \citep[see Figure \ref{fig:all_lc} of this work and Figure 1 from][]{Jiang2022}. After including Swift X-ray and UV observations from late 2021, \cite{Jiang2022} found that the period of the oscillations rapidly decreased from $\sim$1 year to $\sim$1 month over the course of three years. The authors modeled this decreasing periodicity with a system of two merging SMBHs and suggested that they could merge within three years. 

AT2019cuk may present a unique opportunity to potentially witness the coalescence of a binary SMBH for the first time. Hence, there has been extensive multi-wavelength follow-up of AT2019cuk, including new optical \citep[e.g.][]{Moiseev2022,Dotti2023}, radio \citep[e.g.][]{An2022a,An2022b,Bruni2022}, and X-ray observations \citep[e.g.][]{Pasham2022,Dou2022}. Continued optical observations have revealed an evolving spectrum, with the potential appearance of a He I emission line at 6678 \AA\, \citep{Moiseev2022}. \cite{Dotti2023} also present further optical photometry of AT2019cuk and argue against a binary SMBH model based on the decreasing amplitude of the oscillations. Instead, they suggest that the periodicity we see could be the result of Lens-Thirring precession of the accretion disk around only one SMBH.  In the radio, milliarcsecond-resolution very long baseline interferometry revealed a compact ($<0.8$ pc) radio core with a flat radio spectrum producing $\sim$40\% of the radio emission in AT2019cuk, which was suggested to be the result of either an optically-thick jet or corona \citep{An2022b}. However, so far, no radio outburst has been reported, as all reported flux levels are consistent with previous survey observations. Finally, X-ray observations offer unique insight into the inner accretion flow in this source, as, if AT2019cuk is indeed a binary SMBH, then the two putative SMBHs should be too close to host their own broad line regions, but could have their own mini accretion disks \citep[e.g.][]{Shen2010,d'Ascoli2018}. Extensive X-ray follow-up has been reported by \cite{Pasham2022} and \cite{Dou2022}, which both present X-ray spectral analysis of initial NICER, XMM-Newton, NuSTAR, and Chandra observations of AT2019cuk, finding a potentially variable warm absorber and evidence for relativistic reflection.

In this work, we present high-cadence NICER X-ray observations and further optical monitoring of AT2019cuk. When \cite{Jiang2022} first announced AT2019cuk as a potential SMBH binary, we triggered a NICER guest observer program (ID: 4078, PI: Pasham) to perform high-cadence (two visits per day for several months) X-ray monitoring of AT2019cuk starting on January 20, 2022. This monitoring continued until the source reached solar exclusion in late August 2022. In this work, we present results up until September 8, 2022, but note that much of the NICER data is affected by the small sun angle beyond mid August 2022. We also carried out daily optical monitoring using multiple telescopes across the globe. In Section \ref{sec:data}, we describe our data and their respective reduction methods. We present the spectral and timing studies in Section \ref{sec:results} and discuss the implications of our findings on the nature of the X-ray corona and with regards to the potential binary SMBH scenario in Section \ref{sec:disc}. Finally, we summarize our findings in Section \ref{sec:concl}. Throughout this paper, we assume a standard flat $\Lambda$CDM cosmology with $\Omega_\Lambda = 0.73$ and $H_0$ = 70~km~s$^{-1}$~Mpc$^{-1}$. All uncertainties are quoted at the 90\% confidence level, unless otherwise noted.

\section{Observations and Data Reduction}\label{sec:data}

\subsection{NICER} \label{subsec:nicer}

NICER \citep{Gendreau2016} has performed extensive high-cadence monitoring of AT2019cuk, with roughly 1-2 visits per day from January-August 2022. We reduced the NICER data using NICERDAS (version 9, as part of HEASoft version 6.30.1). We first ran \texttt{nicerl2} to reprocess the data with standard NICERDAS tools and apply the default settings, with the following exceptions. We did not perform any undershoot or overshoot filtering (i.e. we use \texttt{underonly\_range=*-*}, \texttt{overonly\_range=*-*}, \texttt{overonly\_expr=NONE}), and instead filtered on the background-subtracted products, using the 3C50 background detailed in \cite{Remillard2022}. We utilized level 3 filtering, which implements a 0.2-0.3 keV background-subtracted count rate threshold of 2 cts s$^{-1}$ and a 13-15 keV background-subtracted count rate threshold of 0.05 cts s$^{-1}$. Any good time intervals (GTIs) which exceed these values were removed from our analysis. Such filtering is designed to limit contamination of the energy range 0.3-12~keV, intended for investigations of the science target. In addition, we also filtered on the cut-off rigidity (COR) parameter in NICER data by using \texttt{cor\_range=1.5-*} when running \texttt{nicerl2} to ensure that any flares from either electron precipitation events in the geomagnetosphere or cosmic rays are removed from our data. We excluded data from detectors 14 and 34, which are known to be excessively noisy, as well as data from any ``hot detectors," defined as a detector whose 0-0.2 keV raw count rate exceeds the median 0-0.2 keV raw count rate for all detectors by $4\sigma$ using an iterative sigma-clipping procedure. Lastly, we looked at the distribution of the 15-18 keV raw count rates, which should all be background events, and removed any GTI which exceeds the median rate for all GTIs by $4\sigma$ using an iterative sigma-clipping procedure.

After all filtering is complete, we extracted a spectrum and estimated the background spectrum for each GTI using the \texttt{nibackgen3C50} tool \citep{Remillard2022}. We utilized the \texttt{nicerarf} and \texttt{nicerrmf} tools to produce response files, with the appropriate weighting for any detectors we removed in filtering. We then grouped each spectrum using the optimal binning procedure \citep{Kaastra2016} and to a minimum of 25 counts per bin to employ $\chi^2$ statistics when fitting. After this grouping, any spectrum with five or fewer channels was discarded as this leaves too few degrees of freedom to perform a fit to a spectral model including both a power law and a blackbody. We included data from January 20, 2022 when the NICER campaign began, up until September 08, 2022 at which point the source was in solar exclusion (ObsIDs 4202540102-108, 4578010101-299). However, much of the data from mid August onward was negatively affected by the small Sun angle. Hence, there is very little reliable data after 3C50 level 3 filtering in ObsIDs greater than 4578010280, although we include any GTI which passes all of the above described cuts. In total, this filtering resulted in 576 GTIs in our analysis for a total exposure time of 323.9~ks.

As \cite{Dou2022} first noted, there is a nearby AGN that is roughly 3.1 arcmin away from AT2019cuk, which is just within the field of view of NICER. Therefore, in Appendix \ref{sec:interloper}, we perform extensive checks to assess the impact of this nearby source. We utilize substantial contemporaneous observations with Swift to assess the impact of the nearby AGN on the NICER light curve, as Swift can separately image each of these sources. No other X-ray sources of comparable flux were detected near AT2019cuk in around 160~ks of stacked Swift data. The NICER light curve closely resembles the Swift light curve for AT2019cuk (see Figure \ref{fig:all_lc} below and Figure \ref{fig:19cuk_swift+nicer} in the Appendix), and Swift also sees occasional hard X-ray spectra (see Figure \ref{fig:gamma_swift+nicer} in the Appendix). Additionally, the nearby AGN's contribution to the NICER measurements is further diluted by a factor of roughly 2-3 due to the reduced effective area of the NICER optics at the offset position (see Figure \ref{fig:arfweight} in the Appendix). Thus, we conclude that the interloper in the NICER field of view contributes negligibly to NICER light curve and spectral analysis.

\subsection{XMM-Newton} \label{subsec:xmm}

XMM-Newton \citep{Jansen2001} performed two DDT observations of AT2019cuk in December 2021 and January 2022 (ObsIDs 0893810201 and 0893810401), which we use for comparison with the NICER data. We reduced the XMM-Newton EPIC-pn data from these two observations using the XMM-Newton Science Analysis System (XMM-SAS; v20.0.0). We followed standard data reduction processes, including running \texttt{epproc} to produce calibrated event files, screening for periods of high background flaring by removing time intervals with a count rate of $>$0.4~cts~s$^{-1}$ in the 10-12~keV range, and producing response files using \texttt{rmfgen} and \texttt{arfgen}. All single and double events (PATTERN $\leq$ 4) were used when extracting spectra. Source spectra and light curves were extracted in a circular region with a radius of 35", and background products were extracted from an off-source region on the same CCD chip with a radius of 40". Spectra were grouped to a minimum of 25 counts per bin to employ $\chi^2$ statistics when fitting.

\subsection{NuSTAR} \label{subsec:nustar}

In February 2022, following the initial announcement of AT2019cuk as a potential SMBH binary, we requested a 100~ks NuSTAR DDT observation \citep[][ObsID 90801605002]{Harrison2013}. All data was processed with NuSTARDAS (v2.1.2, as part of HEASoft v6.30.1) and with calibration files from NuSTAR CALDB v20220802. We reduced the NuSTAR data using \texttt{nupipeline} to produce calibrated event files and \texttt{nuproducts} to produce spectra and light curves for the FPMA and FPMB modules individually. Source products were extracted in a circular aperture with a radius of 60" and background products were extracted in a nearby off-source region with a radius of 80". Spectra were grouped to a minimum of 25 counts per bin to employ $\chi^2$ statistics when fitting.

\subsection{Swift} \label{subsec:swift}

AT2019cuk has also been observed extensively with Swift \citep{Gehrels2004}, with a cadence of roughly one visit every two days from January-August 2022. We utilized the Swift XRT data to provide consistency checks for our NICER analysis and confirm that the NICER light curve is dominated by AT2019cuk, rather than a nearby AGN (see Appendix Section \ref{sec:interloper}). Using the online Swift XRT products generator\footnote{\url{https://www.swift.ac.uk/user_objects/}} \citep{Evans2007,Evans2009}, we created light curves on a per-GTI basis in the same energy bands as NICER (Total: 0.3-4 keV, Soft: 0.3-2 keV, Hard: 2-4 keV). We also extracted spectra on a per-ObsID basis and binned these spectra with the optimal binning procedure \citep{Kaastra2016} and to a minimum of 1 count per bin for the use of $C$ statistic minimization when fitting. This binning was used due to significantly lower count rates from Swift compared to NICER.

During this monitoring, the Swift Ultra-Violet/Optical Telescope (UVOT; \citealt{Roming2005}) instrument performed observations with UVW1 as the primary filter, with only a few observations with all of the other Swift UVOT filters taken at the beginning of the observing period. The data was processed with the \texttt{uvotsource} command. A circular region of 5\arcsec, centered at the target position, was chosen for the source and another region of 40\arcsec~located at a nearby position free of point sources was used to estimate the background emission. UVW1 magnitudes were corrected for Galactic extinction using $E(B-V) = 0.025$ \citep{Schlafly2011}.

\subsection{Ground-Based Optical Photometry} \label{subsec:optphot}

In addition to X-ray and UV monitoring, ZTF has continued to monitor AT2019cuk. We performed point spread function (PSF) photometry on all publicly available ZTF data at the location of AT2019cuk using the ZTF forced-photometry service \citep{Masci2019} in $g$- and $r$-band. Similar to UVOT, ZTF photometry was corrected for Galactic extinction. 

We also collected more optical monitoring from observatories across the globe, including Maidanak Observatory (MAO, 1.5m, \citealt{Ehgamberdiev2018}), Lemmonsan Optical Astronomy Observatory (LOAO, 1.0m), and Chungbuk National University Observatory (CBNUO, 0.6m), all of which are a part of the SomangNet network of telescopes \citep{Im2021}. Both $B$ and $R$ images from the SomangNet telescopes were obtained from February 23, 2022 to July 23, 2022 with roughly daily cadence. The data were reduced using our automatic image analysis pipeline, \texttt{gpPy} (G. S.-H. Paek et al., in preparation), which process includes bias and dark subtractions, flat-fielding, astrometry, and photometry calibration using point sources in the vicinity of AT2019cuk. We performed aperture photometry with a single aperture size to minimize the host galaxy contamination after convolving images from different epochs to have a common point-spread-function (PSF) using the \texttt{HOTPANTS} software (\citealt{Becker2015}; see e.g. \citealt{Kim2018,Kim2019}). Image convolution is necessary due to different seeing values on different nights, which may sample different amounts of the host galaxy light (when done with a variable aperture size that is proportional to the PSF size) or the nuclear transient (when done with a fixed aperture size; see e.g. \citealt{Kim2018}). This process was done independently for each filter and each observatory and means that the optical light curve is contaminated by the host galaxy at different amounts. However, this contamination is negligible since the nuclear component dominates near the center where the aperture photometry was done. As with ZTF and Swift UVOT data, the optical data from the SomangNet network was corrected for Galactic extinction.

\section{Analysis and Results} \label{sec:results}

\begin{figure}[t!]
    \centering
    \includegraphics[width=8.5cm]{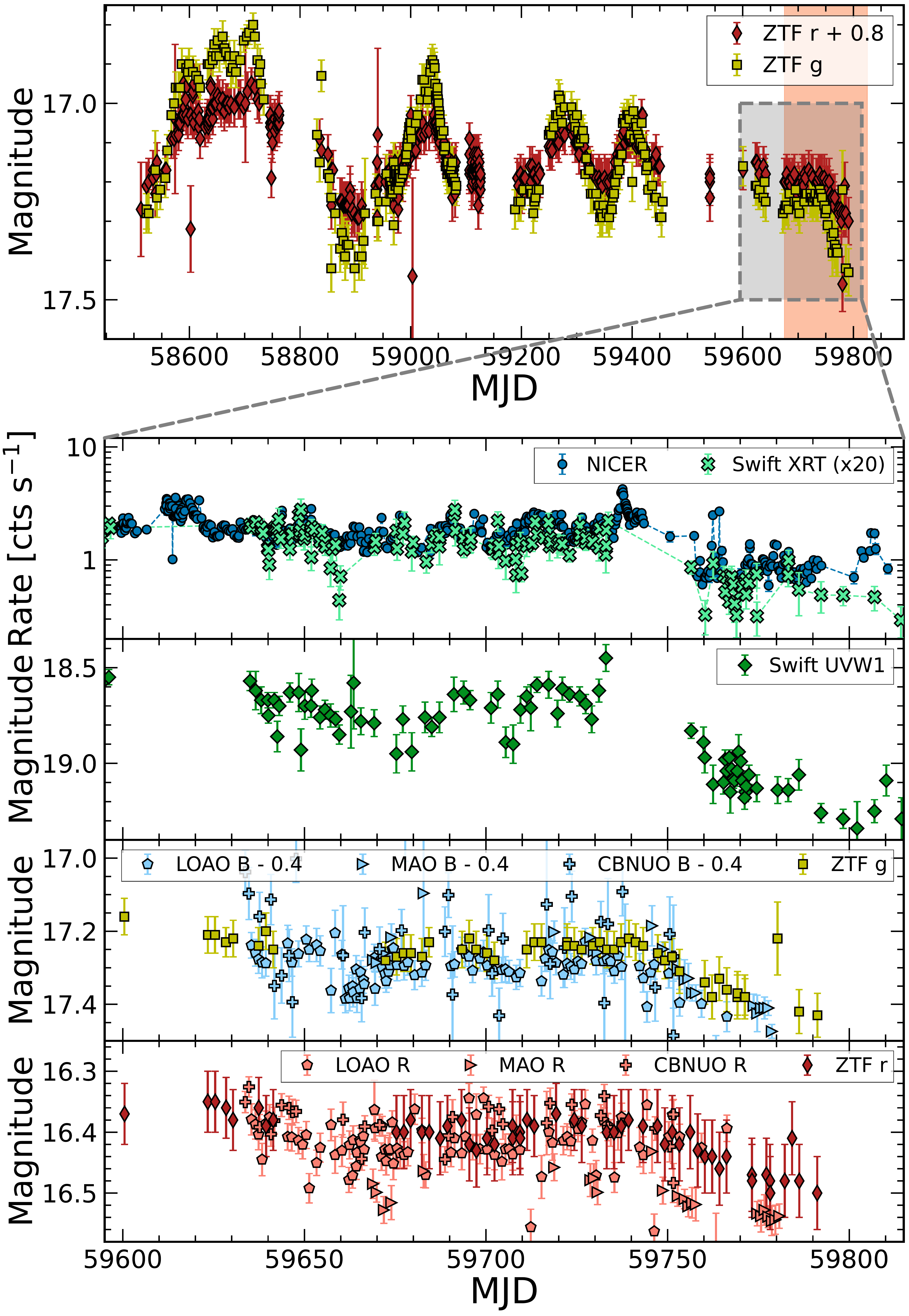}
    \caption{Multi-wavelength light curve for AT2019cuk, with zoom in on the duration of the high-cadence NICER monitoring campaign. \textit{Top panel:} Full ZTF light curve from January 2019 to September 2022. Red diamonds show the ZTF $r$-band data, offset by 0.8 mag, and yellow squares show the ZTF $g$-band data. The orange shaded region shows the prediction of the merger time from \cite{Jiang2022}, using post-Newtonian modeling of both optical and X-ray data \citep[see Figure 3/Extended Figure 9 of][]{Jiang2022}. The grey box represents the zoom in region shown in the lower four panels. \textit{Second panel:} NICER (blue circles) and Swift XRT (green X's) background-subtracted light curves in the 0.3-4~keV band. Swift data are scaled by a factor of 20 to account for the difference in effective area between the two telescopes. \textit{Third panel:} Swift UVW1 light curve. Data have been corrected for Galactic extinction. \textit{Bottom two panels:} Optical light curves, with the fourth panel showing the ZTF $g$-band (yellow squares) and LOAO, MAO, and CBNUO $B$-band (light blue pentagons, triangles, and plus signs respectively) data and the fifth panel showing ZTF $r$-band (red diamonds) LOAO, MAO, and CBNUO $R$-band (light pink pentagons, triangles, and plus signs respectively). There is a $\approx$ 9-10\% drop in the $g$- and $B$-band data in the fourth panel around MJD 59750 and a $\approx$ 3-5\% drop in the $r$- and $R$-band data at the same time, which corresponds to the time at which a large drop in the X-ray flux ($\approx$ 50\%) is also seen.}
    \label{fig:all_lc}
\end{figure}

Figure \ref{fig:all_lc} shows the long-term ZTF light curve spanning from 2019-2022, with a zoom-in on the observing period covered in this paper. The data shown include NICER and Swift XRT in the X-ray (top panel of the bottom zoom-in), Swift UVW1 in the UV (second panel of the bottom zoom-in), and ZTF, CBNUO, MAO, and LOAO in the optical (bottom two panels of the bottom zoom-in). 

\subsection{Identifying \& Validating Hard X-ray Flares} \label{subsec:flares}

\begin{figure*}[t]
    \centering
    \includegraphics[width=18cm]{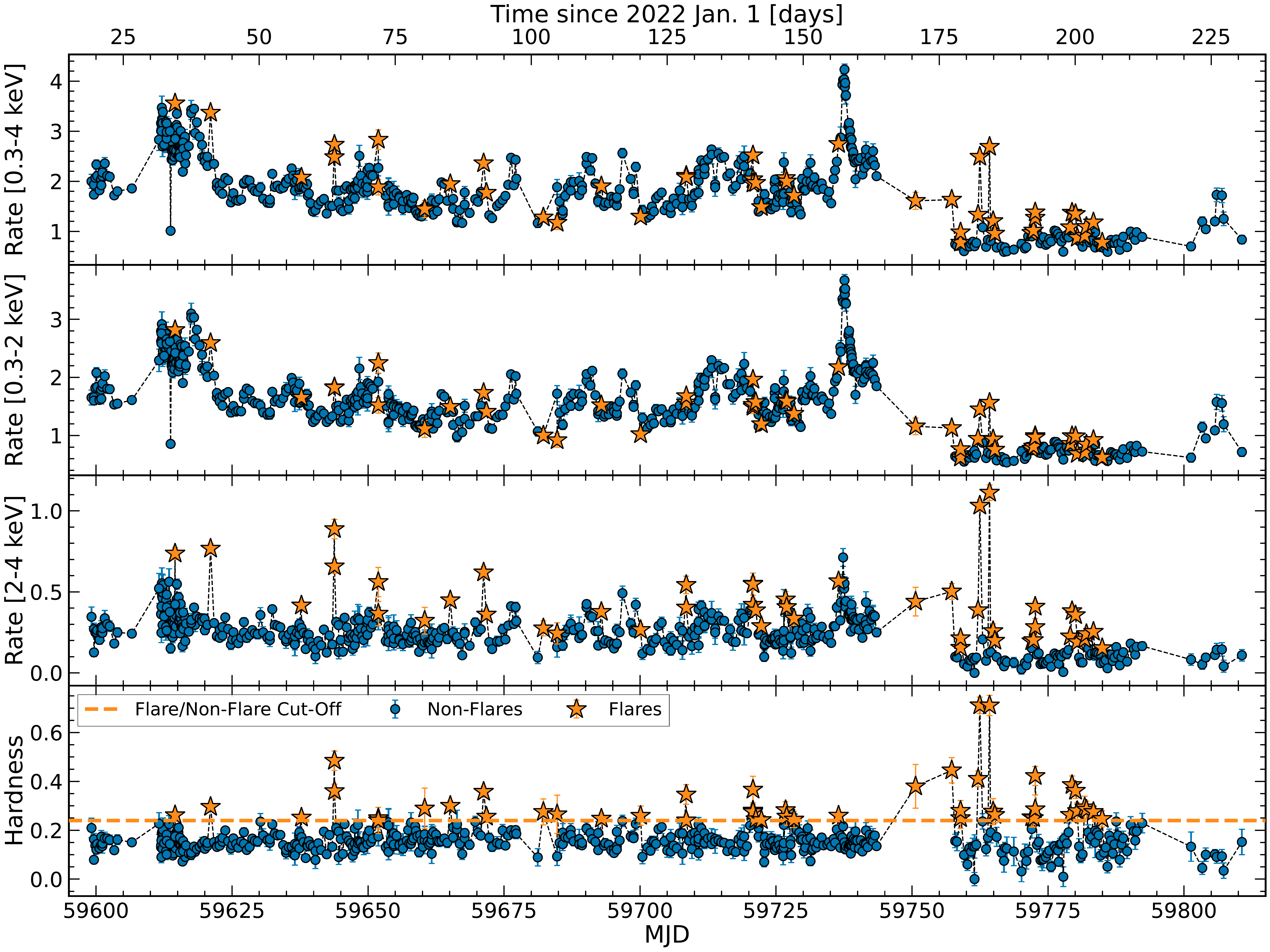}
    \caption{\textit{Top 3 panels:} NICER light curves in three different energy bands (from top to bottom: 0.3-4 keV, 0.3-2 keV, and 2-4 keV) for AT2019cuk. Each light curve is background-subtracted, rescaled to the rate for 52 detectors, and shown in counts per second. \textit{Bottom panel:} NICER hardness ratio, defined as the count rate in 2-4 keV divided by the count rate in 0.3-2 keV. The orange dashed line shows the boundary between our definition of flaring vs. non-flaring GTIs, set at a hardness ratio of 0.2. In all four panels, all dark blue circles correspond to non-flaring GTIs and orange stars correspond to flaring GTIs.}
    \label{fig:picking_flares}
\end{figure*}

One of the most striking features of the AT2019cuk NICER light curve is short, hard X-ray flares, seen most prominently in the 2-4 keV band. In Figure \ref{fig:picking_flares}, we show the NICER light curves for AT2019cuk in three different bands (total: 0.3-4 keV, soft: 0.3-2 keV, hard: 2-4 keV), as well as a hardness ratio, defined as the 2-4 keV background-subtracted count rate divided by the 0.3-2 keV background-subtracted count rate. While the flares are evident by eye in the 2-4 keV light curve and hardness ratio, the GTIs are relatively short and at a low count rate. Thus, great care must be taken to ensure that the flares are indeed real, rather than an artifact of poor counting statistics or contamination from the NICER instrumental effects.

To assess the impact of counting statistics, we used the best fitting XMM-Newton/NuSTAR spectral model to simulate a NICER data set with the same distribution of exposure times and fluxes as the real data. The simulated data show essentially no hard X-ray flares and the real data show a tail of GTIs at significantly higher hardness ratio, indicating that the flares are not the result of poor counting statistics (see Figure \ref{fig:sim_nicer} in the Appendix). Further details of these simulations are given in Section \ref{subsec:verify_sims} of the Appendix. Additionally, as the NICER instrument is most sensitive at soft energies ($\lesssim 2$ keV) and can have a significant background contribution, we are careful to ensure that these flares are astrophysical. We find no evidence for any instrumental or background effect that could lead to such flares, and we detail these extensive checks in Section \ref{sec:verify_nicer} of the Appendix. Similarly, we find consistency between our NICER findings and the contemporaneous Swift observations, including a few similarly hard Swift X-ray spectra (see Section \ref{sec:verify_swift} of the Appendix). Thus, we find that neither counting statistics nor instrumental effects can account for the hard X-ray flares in AT2019cuk.

\begin{figure}[t]
    \centering
    \includegraphics[width=0.47\textwidth]{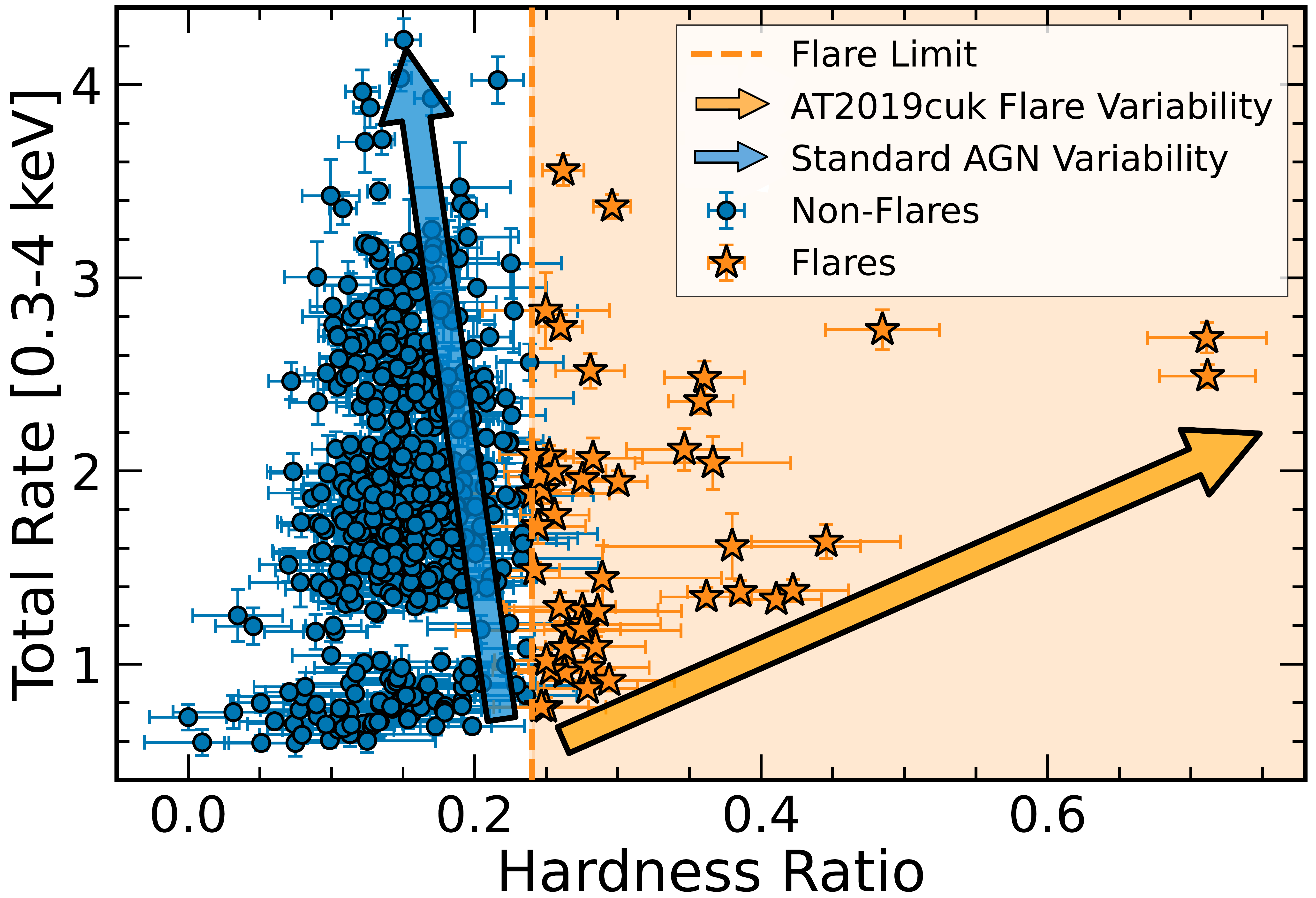}
    \caption{Hardness-intensity diagram for the NICER data. Flaring GTIs are shown as orange stars (any point with a hardness ratio greater than 0.24), and non-flaring GTIs are shown as blue circles. The blue arrow shows a representative track for the standard softer-when-brighter behavior seen in AGN \citep[e.g.][]{Sobolewska2009}, which is similar to what is seen in the non-flaring points for AT2019cuk. The orange arrow shows the harder-when-brighter behavior seen in the hard X-ray flares of AT2019cuk, which are starkly different from the standard AGN variability that is expected.}
    \label{fig:hid}
\end{figure}

Identifying flares is best done using the hardness ratio, as it normalizes the long term variability using the soft band. We again utilize the simulated NICER data set to help identify flares, as it provides a less biased distribution with which to distinguish outliers. Therefore, we define a flare to be any GTI with a hardness ratio greater than $3\sigma$ above the mean hardness ratio of the simulated NICER data (which corresponds to a hardness ratio cut of 0.24). These flares are shown as orange stars in Figure \ref{fig:picking_flares} and throughout the rest of this paper. As evidenced by the third panel of Figure \ref{fig:picking_flares}, this hardness ratio cut picks out the hard flares in the 2-4 keV band quite well. Additionally, we show a hardness-intensity diagram in Figure \ref{fig:hid}, which highlights the unique variability properties of the flares in AT2019cuk. For comparison, Figure \ref{fig:hid} shows a blue arrow highlighting the expected coronal variability for AGN which follow the standard softer-when-brighter behavior \citep[e.g.][]{Sobolewska2009}. The hard X-ray flares show variability that is opposite of what is usually seen in AGN.


Given the hard X-ray nature of these flares, we searched the Swift BAT 105-month catalog \citep{Oh2018} and the Swift BAT transient monitoring catalog \citep{Krimm2013} for AT2019cuk, but the source was not detected. Given the transient nature of this source and the relatively low X-ray count rate, it is not particularly surprising that it was not detected by Swift BAT, given the relatively high flux sensitivity limits of Swift BAT.

\subsection{Timing Properties of the Hard X-ray Flares} \label{subsec:x-ray_timing}

The average duration of a flare is roughly 1~day, with measurements ranging from 0.9-1.6~days depending on how the flare times are measured and whether we include isolated flares (i.e. flares that last for only a single GTI). The average time between flares is roughly 5~days, as measured from the last GTI of one flare to the first GTI of the next flare. However, we note that the spread of the time between flares ranges from approximately 1~day to up to 15~days and is much larger than that of the flare duration. It is, however, worth noting that there is significant variability even within a given NICER GTI for AT2019cuk, as the longest GTIs (with more than 1000 seconds of exposure) tend to show a factor of two change in hardness on timescales as short as 200 seconds. Despite the repetitive nature of the flares, a Lomb-Scargle analysis on either the hard X-ray band (2-4 keV) or the hardness ratio over the entire observing period shows no significant periodicity or quasi-periodicity.

\subsection{X-ray Spectral Modeling} \label{subsec:spec}

All spectral fitting detailed in this section is performed in XSPEC \citep[v12.12.0;][]{Arnaud1996}.

\subsubsection{XMM-Newton/NuSTAR} \label{subsubsec:spec_xmm}

We utilize the high quality XMM-Newton and NuSTAR spectra to perform spectral modeling that informs how we fit the short NICER observations. We simultaneously fit the two XMM-Newton spectra from 0.3-10~keV and the two spectra from NuSTAR FPMA and FPMB from 3-70~keV. Each redshifted model component is fixed at $z = 0.081$, abundances are adopted from \cite{Wilms2000}, and cross-sections are adopted from \cite{Verner1996}. The column density of Galactic absorption (\texttt{tbabs} component in XSPEC) is fixed at the HI value of $N_H = 2.26 \times 10^{20}$ cm$^{-2}$ \citep{HI4PICollaboration2016}. We allow for a constant offset between the XMM-Newton spectra and each NuSTAR spectrum for cross-calibration and to account for the lack of simultaneity of these observations. Otherwise, all parameters are tied across all four spectra when fitting. 

\begin{figure}[t]
    \centering
    \includegraphics[width=0.47\textwidth]{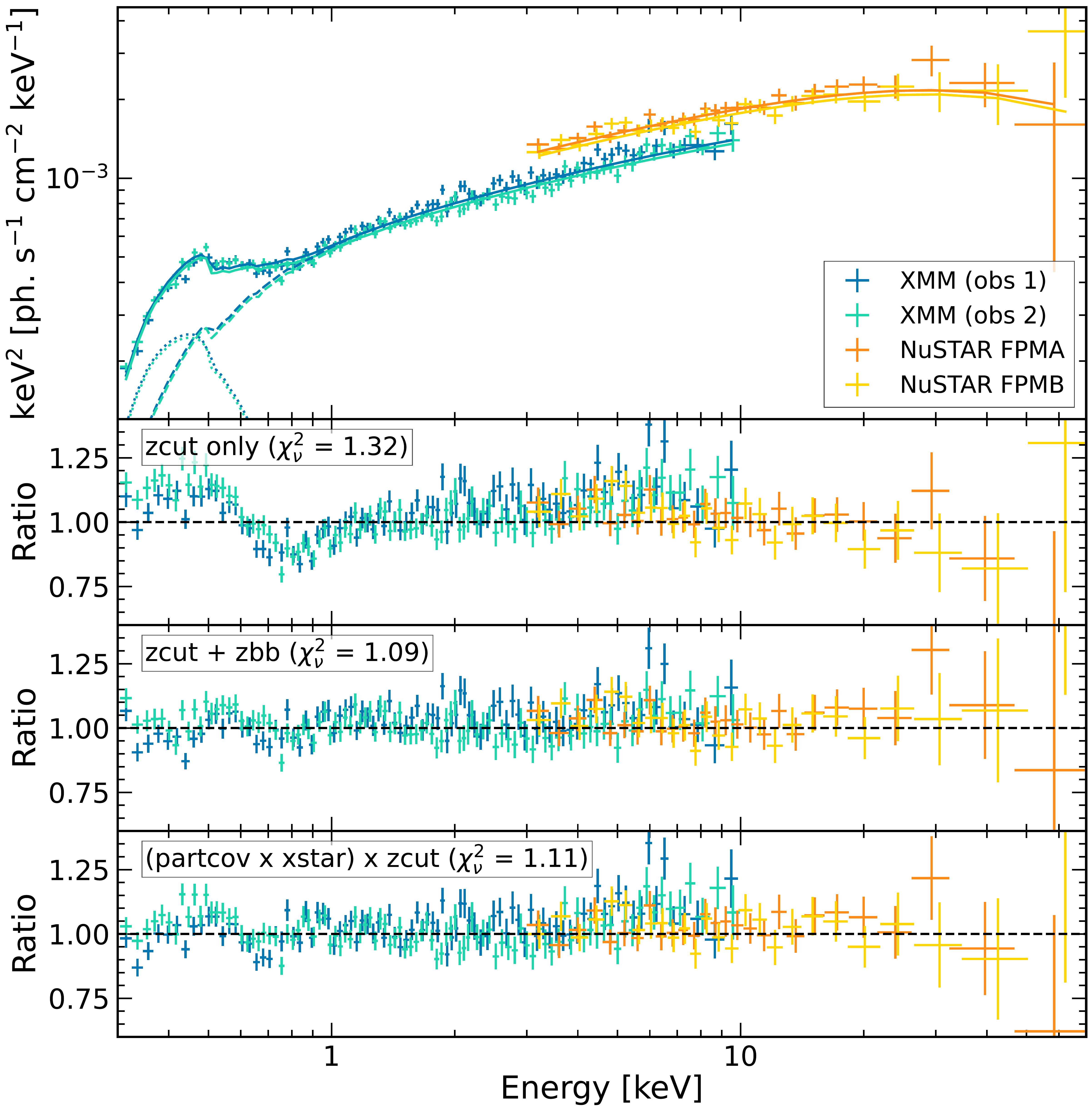}
    \caption{Spectral fits to the XMM-Newton observations from December 2021 (Obs. 1, shown in light green) and January 2022 (Obs. 2, shown in dark blue) and the NuSTAR observation from January 2022 (FPMA/FPMB shown in dark/light orange, respectively). \textit{Top panel:} Unfolded spectrum fit to the \texttt{tbabs*ztbabs*(zcut+zbb)} model. \textit{Second panel:} Ratio plot for the \texttt{tbabs*ztbabs*zcut} model. Clear residuals are seen in the soft part of the spectrum. \textit{Third panel:} Ratio plot for the \texttt{tbabs*ztbabs*(zcut+zbb)} model. \textit{Bottom panel:} Ratio plot for the \texttt{tbabs*(partcov*xstar)*zcut} model. Both the partial covering ionized absorber and the blackbody give statistically comparable fits and significant improvement over the cutoff power law alone.}
    \label{fig:nuxmm}
\end{figure}

A simple absorbed cutoff power law (i.e. \texttt{tbabs*ztbabs*zcut} in XSPEC notation) produces a poor spectral fit ($\chi^2_\nu = \chi^2/\mathrm{dof} = 1.32$) with significant residuals in the soft part of the spectrum (see second panel of Figure \ref{fig:nuxmm}). Adding an additional blackbody as a phenomenological model for the soft excess (i.e. \texttt{tbabs*ztbabs*(zcut+zbb)} in XSPEC notation) improves the fit and provides a statistically good fit to the data ($\chi^2_\nu = 1.09$). As a more physically-motivated approach, we also explored whether the soft features in the XMM-Newton spectra can be modeled with a warm absorber \citep[similar to the modeling presented in][]{Dou2022}. To model this warm absorber, we produced a grid of photoionization models with variable column density and ionization parameter using \textsc{xstar} \citep{Kallman2001}, assuming a power law ionizing spectrum with $\Gamma = 2$ and a turbulent velocity of $v_\mathrm{turb} = 300$~km~s$^{-1}$. This ionized absorption fit is statistically comparable to the \texttt{zcut+zbb} model with $\chi^2_\nu = 1.11$, with no need for an additional neutral absorber or the blackbody component (i.e. \texttt{tbabs*(partcov*xstar)*zcut} model in XSPEC). In this model, we leave the column density, ionization parameter, covering fraction, and redshift of the ionized absorber free when fitting, but find that at 90\% confidence, the absorber is consistent with being at the redshift of the source. The blueshift of the absorber could be determined with the higher energy resolution of the RGS instrument on XMM-Newton, but the low flux in these observations makes the RGS observations very close to the background level. 

The fit results for each of these models are given in Table \ref{tab:xmm+nu}, and a ratio plot for each of these fits is shown in Figure \ref{fig:nuxmm}. Overall, the XMM-Newton and NuSTAR data reveal that in early 2022 AT2019cuk had a fairly normal X-ray spectrum compared to other AGN. In particular, the photon index and cutoff energy are comparable to standard AGN \citep[e.g.][]{Ricci2017}, and the ionized absorber is similar to the commonly-observed warm absorber in AGN. 

\subsubsection{NICER} \label{subsubsec:spec_nicer}

To assess the spectral evolution during the hard X-ray flares in AT2019cuk during the NICER observing campaign, we perform spectral modeling of the NICER data on a per-GTI basis. We apply the same models that we fit to XMM-Newton/NuSTAR in the previous section, except that we utilize a power law rather than a cutoff power law when fitting NICER data as the NICER instrument has lower effective area at high energies and becomes background-dominated at much lower energy than we have access to with NuSTAR. All fits with NICER are performed in the 0.3-4~keV range, although we also fit each GTI in the 0.3-$E_\mathrm{upper}$~keV range, where $E_\mathrm{upper}$ denotes the energy at which the background begins dominating over the source emission, and found that there is no considerable difference to the fit results. However, at late times (MJD $>$ 59750) when the sun angle is decreasing and the overall flux has dropped, some of the GTIs are dominated by the background below 1~keV, which makes it very difficult to properly constrain fit parameters like $\Gamma$. Hence, we exclude the GTIs with $E_\mathrm{upper} < 1$~keV in spectral fitting. Additionally, for each model, around 5\% of GTIs are poorly fit with $\chi^2_\nu > 2$, which also primarily occurred at late times when the soft part of the spectrum was potentially contaminated by the NICER noise peak due to low sung angle and the baseline model for AT2019cuk (set by the XMM-Newton/NuSTAR data from December 2021/January 2022) may have changed. Thus, we excluded these GTIs from our spectral analysis. When included, we bound the column density of the neutral absorber to be larger than $N_H = 10^{19}$~cm$^{-2}$, the column density of the ionized absorber to be within $N_H = 10^{21}-10^{24}$~cm$^{-2}$, the blackbody temperature to be within $kT_\mathrm{bb} = 0.01-0.3$~keV, and the photon index to be within $\Gamma = 0-3$. We also remove GTIs in which either the blackbody temperature, the photon index, or the column density of the ionized absorber was unconstrained between the two limits provided. 

\begin{deluxetable}{c c c}
	\caption{XMM-Newton/NuSTAR Spectral Fits} \label{tab:xmm+nu}
    \tablehead{\colhead{Model} & \colhead{Parameter (unit)} & \colhead{Value}}
	\startdata
	\texttt{zcut} & $\Gamma$ & $1.65_{-0.01}^{+0.01}$ \\
    & $E_\mathrm{cut}$ (keV) & $>173$ \\
    \texttt{zcut+zbb} & $N_{H,\mathrm{neut}}$ ($10^{22}$ cm$^{-2}$) & $0.07_{-0.01}^{+0.01}$ \\
     & $\Gamma$ & $1.59_{-0.03}^{+0.03}$ \\
     & $E_\mathrm{cut}$ (keV) & $75_{-18}^{+33}$ \\
     & $kT_\mathrm{bb}$ (eV) & $81_{-3}^{+3}$ \\
     \texttt{(partcov*xstar)*zcut} & $N_{H,\mathrm{ion}}$ ($10^{22}$ cm$^{-2}$) & $1.1_{-0.2}^{+1.4}$ \\
     & $\log(\xi/\mathrm{erg}\,\mathrm{cm}\,\mathrm{s}^{-1})$ & $1.91_{-0.14}^{+0.11}$ \\
     & Covering Fraction & $0.88_{-0.36}^{+0.12}$ \\
     & $\Gamma$ & $1.73_{-0.01}^{+0.01}$ \\
     & $E_\mathrm{cut}$ (keV) & $>281$ \\
    \enddata
\end{deluxetable}

When modelled phenomenologically with a blackbody and power law (i.e. \texttt{tbabs*ztbabs*(zpo+zbb)} in XSPEC notation), the largest difference between the flaring and non-flaring spectra is the distribution of photon indices. The weighted mean of the flaring photon indices is $\langle\Gamma_\mathrm{flare}\rangle = 0.77 \pm 0.17$ (where the error represents the $1\sigma$ scatter in the distribution), which is significantly lower than the non-flaring spectra ($\langle\Gamma_\mathrm{non-flare}\rangle = 1.47 \pm 0.25$) and the distribution that is seen in standard AGN \citep[$\Gamma \approx 1.6-2.0$; e.g.][]{Ricci2017}. We also test a partial covering neutral absorber (i.e. \texttt{tbabs*pcfabs*(zpo+zbb)}) with the photon index and blackbody temperature fixed at the values from the XMM-Newton/NuSTAR best-fit model to explore potential degeneracies between the photon index and neutral column density. This model yields comparable fit statistics with changing the photon index, but requires roughly a simultaneous increase in the column density and normalization of the power law, the implications of which we discuss further in Section \ref{subsec:disc_obs}.

\begin{figure}[t]
    \centering
    \includegraphics[width=0.47\textwidth]{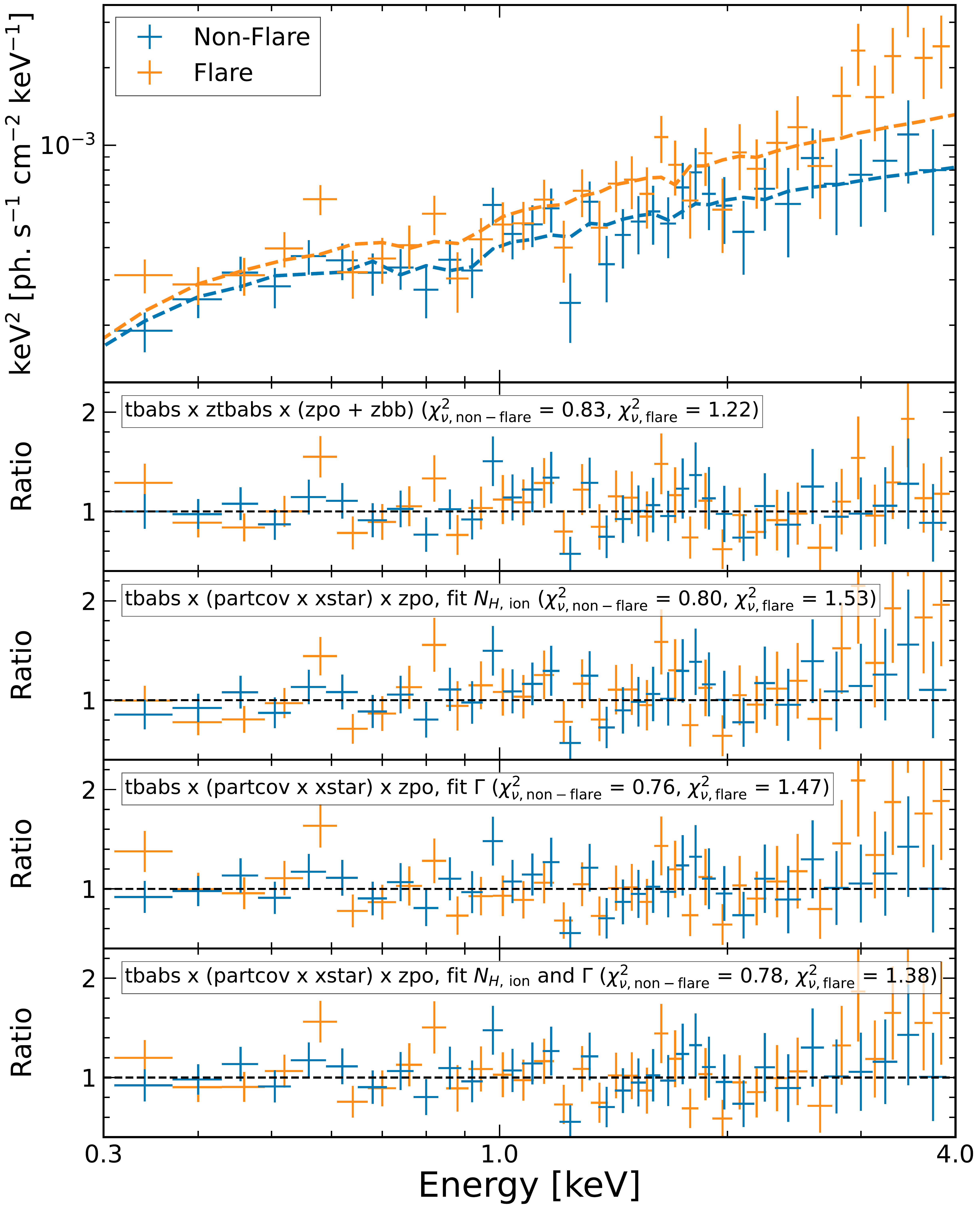}
    \caption{Spectral fits to example non-flaring and flaring NICER GTIs. The non-flaring spectrum is shown in blue, and the flaring spectrum is shown in orange. \textit{Top panel:} Unfolded spectrum fit to the \texttt{tbabs*(partcov*xstar)*zpo} model with the photon index free between the two GTIs. \textit{Second panel:} Ratio plot for the \texttt{tbabs*ztbabs*(zpo+zbb)} model. \textit{Third panel:} Ratio plot for the \texttt{tbabs*(partcov*xstar)*zpo)} model with the column density of the ionized absorber free. The photon index and the ionization parameter and covering fraction of the ionized absorber are fixed at the best fit values from the XMM-Newton/NuSTAR fit. \textit{Fourth panel:} Ratio plot for the \texttt{tbabs*(partcov*xstar)*zpo)} model with the photon index of the power law free. The column density, the ionization parameter, and covering fraction of the ionized absorber are fixed at the best fit values from the XMM-Newton/NuSTAR fit. \textit{Bottom panel:} Ratio plot for the \texttt{tbabs*(partcov*xstar)*zpo} model, but with both the column density of the ionized absorber and the photon index free.}
    \label{fig:flare_vs_nonflare}
\end{figure}

\begin{figure}[t]
    \centering
    \includegraphics[width=0.47\textwidth]{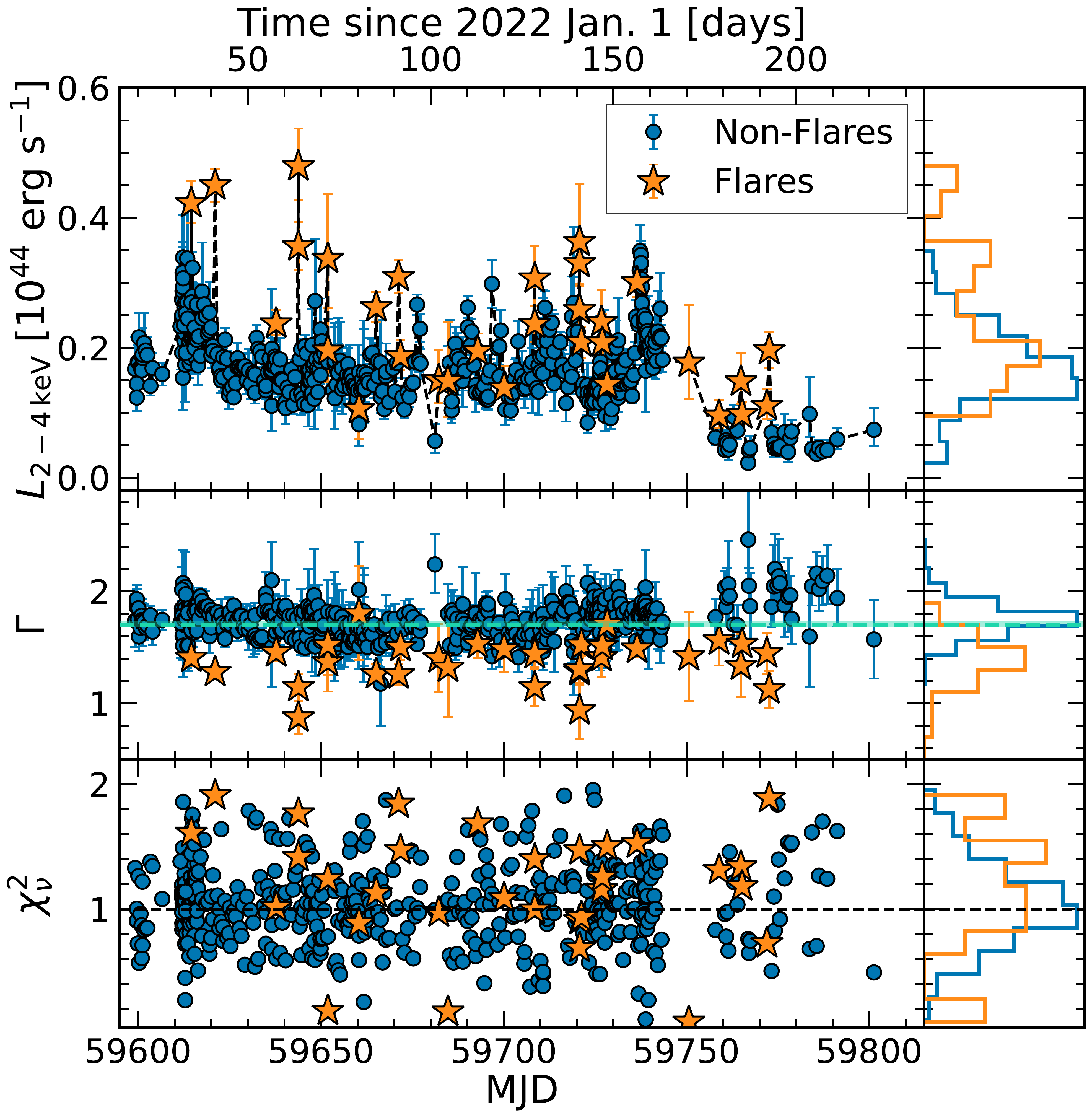}
    \caption{NICER spectral fits to the ionized absorption model with a variable photon index (i.e. \texttt{tbabs*(partcov*xstar)*zpo}). In each panel, the dark blue circles are the non-flaring data and orange stars the flaring data (as defined by the hardness ratio cut detailed in Section \ref{subsec:flares}). The left shows all of the NICER data, with each point corresponding to a single NICER GTI, whereas the right shows a histogram of the flaring and non-flaring data over the entire observing period. \textit{Top Panel:} Hard X-ray luminosity in the 2-4~keV band. \textit{Middle Panel:} Photon index of the power-law component. The horizontal light green dashed line shows the best-fit value for $\Gamma$ from the XMM-Newton/NuSTAR fit, with the shaded region showing the errors. There is generally good consistency between the non-flares and the XMM-Newton/NuSTAR value, but the flaring data have significantly lower photon indices on average. \textit{Bottom Panel:} Reduced $\chi^2$ statistic. Both the flares and the non-flares are similarly well-fit by this model.}
    \label{fig:xstar_nicer}
\end{figure}

We also explore whether this difference between flares and non-flares can be explained by changes in an ionized absorber by fitting each NICER GTI with the ionized absorption model (\texttt{tbabs*(partcov*xstar)*zpo} in XSPEC). Due to the limited spectral quality in individual NICER GTIs, we fix the covering fraction and ionization parameter at the XMM-Newton/NuSTAR values given in Table \ref{tab:xmm+nu}. We then proceed with two possible models---one in which we fix the photon index and fit for the column density of the ionized absorber, and vice versa. We find that these two models give similar fit statistics on average ($\langle\chi^2_{\nu,\,\mathrm{flare}}\rangle \approx 1.14-1.19$), comparable to the values from the phenomenological model previously discussed. With the column density free, both an increase in the intrinsic normalization of the power law and the column density of the ionized absorber are required, similar to the behavior seen in the partial covering neutral absorption model (see Section \ref{subsec:disc_obs} for further discussion). This model produces similar fit statistics on average compared to the free photon index fits, but the free column density model struggles to produce the hardest X-ray spectra seen in the most extreme flares. With the photon index free, the flares have $\langle\Gamma_\mathrm{flare}\rangle = 1.37 \pm 0.19$, which is much more reasonable compared to the phenomenological model in which the blackbody may be leading to artificially low photon indices by trying to reproduce the features of the warm absorber. However, these photon indices are still harder than standard AGN and the non-flares, which have $\langle\Gamma_\mathrm{non-flare}\rangle = 1.75 \pm 0.14$. This intrinsic change to the spectral shape could result from increased magnetic activity in the corona, which we discuss further in Section \ref{subsec:disc_jet}. In Figure \ref{fig:flare_vs_nonflare}, we show an example of one non-flaring GTI (from ObsID 4578010170 on MJD 59672) and one flaring GTI (from ObsID 4578010169 on MJD 59671), along with various model fits. The results of the ionized absorber fits with a free photon index are shown in Figure \ref{fig:xstar_nicer}, which highlights the clear difference between the photon indices in flares and non-flares.

In the case where we allow both the column density of the ionized absorber and the photon index to vary, there are degeneracies between a decreasing photon index and an increasing absorber column density, as both produce harder X-ray spectra, as seen in the flares. These degeneracies are difficult to decouple with the limited spectral quality of single NICER GTIs and limited energy range considered. We do find a statistically significant improvement in the average fit statistic ($\langle\chi^2_{\nu,\,\mathrm{flare}}\rangle = 1.01$) when allowing both parameters to be free, but it is difficult to reconcile a simultaneous increase in the column density and hardening of the coronal emission physically. Similarly, allowing only the ionization parameter to be free, rather than the column density, requires a simultaneous decrease to the ionization parameter and an increase in the normalization (i.e. intrinsic luminosity) of the power law. Such changes are counter-intuitive based on the definition of the ionization parameter ($\xi = L_\mathrm{ion}/nr^2$), and thus indicate that either a significant increase in density or distance of the obscuring material would be required, neither of which are feasible on such short timescales. 

\subsection{Optical/UV Variability During NICER Coverage} \label{subsec:opt}

The optical/UV light curves are shown in the bottom three panels of Figure \ref{fig:all_lc}, with the full ZTF light curve from 2019 onward shown in the top panel for reference. The most notable feature of the multi-wavelength data is the lack of variability on roughly 30~day timescales or shorter, as seen in and predicted by \cite{Jiang2022}. There is an interesting spike in the X-rays around MJD 59740, followed by a significant decrease in X-ray flux, which is also seen in the UV and optical bands. However, the variability seen in the drop is similar to standard AGN behavior, whereby the strongest variability is in X-rays and decreases with increasing wavelength. This sort of behavior can be explained by the standard reprocessing scenario, whereby X-rays illuminate the disk and imprint their variability on short timescales, which gets dampened by reprocessing at longer wavelengths \citep[e.g.][]{Edelson1996,Cackett2007,Panagiotou2022}.

\section{Discussion} \label{sec:disc}

\subsection{Possible Origins of the Hard X-ray Flares}

\begin{figure*}[t!]
    \centering
    \includegraphics[width=\textwidth]{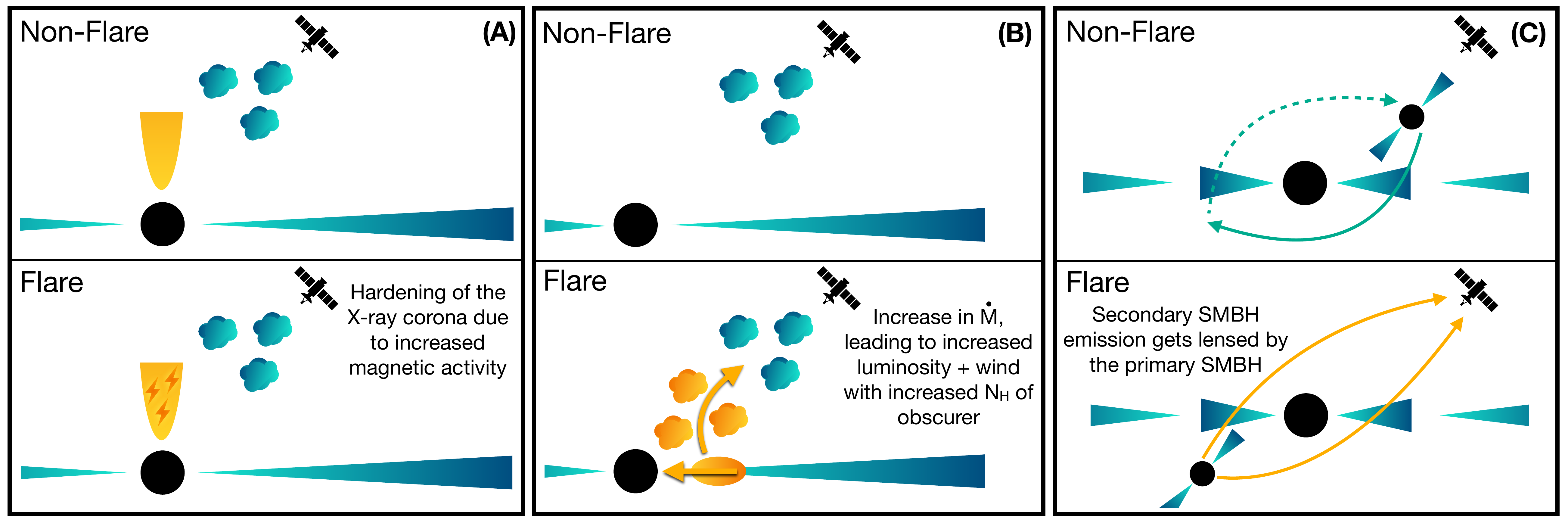}
    \caption{Schematics for potential flare scenarios. \textit{Panel (A):} Flares driven by intrinsic changes in the X-ray corona. Increased magnetic activity in the X-ray corona can lead to magnetic reconnection events causing the observed spectrum to harden. \textit{Panel (B):} Flares driven by changes in the line-of-sight absorption. The flares correspond to an increase in mass accretion rate, which drives a simultaneous increase in the observed luminosity and a wind that increases the observed column density of the absorber. \textit{Panel (C):} Flares are the result of gravitational self-lensing in a binary SMBH.}
    \label{fig:schem}
\end{figure*}

X-ray spectra with photon indices as low as $\Gamma \sim 1$, as seen in the flares of AT2019cuk, are highly unusual for AGN. We propose three potential models for these exotic hard X-ray flares, detailed below and shown schematically in Figure \ref{fig:schem}. 

\subsubsection{A Variable Corona/Jet} \label{subsec:disc_jet}

While the photon indices reached in the flares of AT2019cuk are rather low compared to standard AGN, they do resemble some of the more extreme values seen in some jetted systems like blazars \citep[e.g.][]{Giommi2012,Tagliaferri2015,Bhatta2018} and jetted tidal disruption events \citep[TDEs; e.g.][]{Saxton2012}. Similarly, rapid hard X-ray variability reminiscent of the flaring in AT2019cuk is seen in both blazars \citep[e.g.][]{Hayashida2015,Giommi2021}, jetted TDEs \citep[e.g.][]{Burrows2011,Reis2012}, and black hole binaries \citep[e.g.][]{Walton2017}. In the jetted TDE Swift J1644+57, this hard X-ray variability has previously been attributed to the jet wobbling into and out of our line of sight \citep[see][]{Tchekhovskoy2014}, on timescales on the order of hours to days, similar to the timescale of flares in AT2019cuk. However, AT2019cuk appears to be a radio-quiet AGN, with VLBI observations showing that the radio emission is rather weak (with radio-loudness parameter $R \approx$ 1) and contained within $\lesssim$100~pc \citep{An2022b}. Similarly, the optical data do not show rapid variability and the luminosity of AT2019cuk is too dim to be relativistically beamed, both of which are common properties of blazars. 

Therefore, while we do not expect these flares to arise from a powerful jet pointed along our line of sight, we can draw on the similarity of the flares to the hard X-ray behavior of jetted systems. Magnetic reconnection has long been proposed as a potential heating mechanism for the X-ray corona in AGN and black hole binaries \citep[e.g.][]{Galeev1979} as well as for the acceleration of particles in black hole jets \citep[e.g.][]{Sironi2015,Sironi2020}. Simulations of particle acceleration in jets have shown that magnetic reconnection can produce relatively hard particle spectra ($p \sim 1.2$ in the limit of large magnetization, where the particle distribution is a power law with $n(\gamma) \propto \gamma^{-p}$). If these particles are emitting X-rays through a non-thermal radiative mechanism, then the resulting spectrum would be a power law with photon index $\Gamma = 1 + \frac{p-1}{2} \sim 1.1$ \citep[e.g.][]{Guo2014,Sironi2014}, comparable to what we see in the flares of AT2019cuk. GRMHD simulations suggest that the base of the jet should have high magnetization \citep[e.g.][]{McKinney2006,Tchekhovskoy2011,Chatterjee2019}, and is thus a natural place for magnetic reconnection to occur. Magnetic reconnection is therefore a viable method to produce hard X-ray spectra with non-thermal particle distributions, but such distributions are rather unusual in radio-quiet sources like AT2019cuk, which typically have coronae that produce a thermal Comptonization spectrum and are limited in spectral hardness by pair production.

A hardening of the observed coronal emission can also be explained by a dynamic and outflowing corona, as fewer disk photons reach and cool the corona, leading to extreme heating and photon-starvation of the corona \citep[e.g.][]{Beloborodov1999}. Outflowing coronae have been claimed in a handful of AGN, but as evidenced from low reflection fractions rather than abnormally hard X-ray spectra \cite[e.g.][]{Markoff2004,Markoff2005,Wilkins2015,King2017,Wilkins2022}. Unfortunately, we cannot simultaneously probe the reflection fraction and photon index with the existing NICER data due to the background-dominated nature of the spectrum in the Fe K band. However, the XMM-Newton/NuSTAR observations presented in this work show only a weak Fe K line, suggesting that a low reflection fraction is not unlikely in this system, even in the non-flaring spectrum. 

We suspect that the hardening of the X-ray spectrum during the flares is likely related to increased magnetic activity in the corona driving repetitive reconnection events (see the left panel of \ref{fig:schem}). Despite being more extreme than standard corona variability, the timescales in which these flares occur are not atypical for coronal variability seen in typical AGN. One potential way to test this corona model for the X-ray flares is through high-cadence radio monitoring to probe the variability of the parsec-scale jet, as the jet and corona are likely closely related.

\subsubsection{Variable Obscuration} \label{subsec:disc_obs}

When modeled with a simple phenomenological model such as a power law, highly obscured AGN can exhibit effective photon indices close to $\Gamma_\mathrm{eff} \sim 1$, as a result of not properly accounting for obscuration \citep[e.g.][]{Rovilos2014,Ricci2017,Wang2022}. In general, the degeneracy between the obscuring column density and the photon index can only be broken by going to higher energies ($E > 10$~keV), which are less affected by the obscuring material than softer energies. Thus, with its extended energy range and high effective area relative to other hard X-ray observatories, NuSTAR provides crucial constraints on the nature of highly obscured AGN \citep[e.g.][]{Arevalo2014,Balokovic2014,Bauer2015,Brightman2015,Marchesi2018,LaMassa2019}. The NuSTAR observation of AT2019cuk shows a rather typical X-ray spectrum without significant signs of heavy neutral obscuration. Likewise, both the NuSTAR and XMM-Newton observations lack a strong narrow Fe K$\alpha$ line, which is expected and commonly observed in Compton-thick AGN \citep[e.g.][]{Matt1996,Levenson2006}. Thus, we don't expect to see such heavy obscuration reminiscent of Compton-thick AGN in AT2019cuk, but lower levels of obscuration could still play a role in giving the flares their hard X-ray spectral shape.

Obscuration alone is expected to produce dips in the soft band while the hard band should remain relatively unchanged, since obscuration preferentially attenuates the soft X-ray band. Thus, to obtain the hard X-ray flares seen in AT2019cuk, something more complicated than a simple increase in the amount of obscuration must be going on. One way to increase the hard X-ray flux is to increase in the mass accretion rate, but this must be coupled to an increase in the level of obscuration that leads to a roughly steady soft X-ray flux. Such a scenario could occur if a sudden influx of material drives a wind that increases the amount of obscuring material (see middle panel of Figure \ref{fig:schem}). However, this scenario requires fine-tuning as the timescales for increasing the luminosity and obscuration must be comparable in every flare seen. The day-long timescales in the flares are also extremely fast for a black hole with a mass of roughly $10^8$ $M_\odot$ and are comparable to the dynamical timescale within roughly tens of gravitational radii. The rapid variability in the obscuration would require the wind to be driven from very close to the black hole, which is much closer than where either the neutral or mildly ionized absorbers are typically found in AGN \citep[see e.g.][and references therein]{Laha2021}. 

Another possibility is that the flares could arise from the brief unveiling of a highly obscured sight line. In this scenario, the flares are the result of the column density of this sight line decreasing (down to $N_H \approx 10^{23}$ cm$^{-2}$), temporarily revealing a hard X-ray spectrum. The increase in the observed luminosity could then be due to this additional light reaching us, rather than an intrinsic change to the mass accretion rate. To produce such rapid and repetitive spectral changes, the obscuring material must be in an optically-thick, clumpy torus, and to avoid a strong narrow Fe K$\alpha$ line that is commonly seen in highly obscured AGN, the obscuring material must either have a low covering fraction or be oriented close to edge-on such that the opposite side of the nucleus is obscured. However, this is difficult to reconcile with the fact that the system spends most of its time ($\sim92\%$) in the non-flaring, typical unobscured AGN state.


\subsubsection{Self-Lensing Events in a Binary System} \label{subsec:disc_lens}

Another possible model for the repeating flares in AT2019cuk is self-lensing in a binary SMBH, whereby emission from an accretion flow around one or both SMBHs in a binary can be gravitationally lensed once per orbit and generate repeating flares \citep{D'Orazio2018a}. One potential self-lensing flare was reported in the Kepler light curve of a binary SMBH candidate \citep{Hu2020}. For SMBHs with $M \approx 10^{6-9}~M_{\odot}$ and with orbital periods of days to years, the predicted lensing flares are bright ($10\%-1000\%$ magnification), last days to months in duration, and should occur in roughly 10\% of accreting SMBH binaries \citep{D'Orazio2018a}. 

The self-lensing model predicts that the only wavelength dependence in the flares should arise from a finite source size with respect to the Einstein radius of the lensing SMBH. The Einstein radius scales as $M_\mathrm{BH}^{1/2}$, but the emitting region at a fixed gravitational radius scales linearly with the black hole mass. Thus, finite source size effects are expected in higher mass systems, and in particular are possible for X-ray emission from within $R \approx 10 R_g$ for a $M_\mathrm{BH} \approx 10^8  \, M_\odot$ binary with a roughly equal mass ratio. Thus, these finite source effects could potentially produce the hardening of the X-ray spectrum, if the soft X-rays are produced at a larger radius than the hard X-rays. Additionally, the lack of flares in the UV and optical bands could be a result of most of that emission originating in a circumbinary disk, which is not lensed.

On the other hand, the lack of periodicity in the hard X-ray flares is inconsistent with the expectations of the self-lensing hypothesis, as the flares should arise on timescales set by the orbital period. One possible way around this requirement is with precession, but this would require the recurrence time of the flares to be correlated and no such behavior is seen in AT2019cuk. Hence, the irregular repetition of the flares in AT2019cuk poses a significant problem for the standard self-lensing hypothesis. 

\subsection{No Quasi-Periodic Behavior on Timescales of 30 days}

None of the data collected since early 2022 shows a clear modulation on the order of 30 days or less, as would have been expected for the proposed binary SMBH system \citep{Jiang2022}. The NICER data are potentially contaminated by a nearby AGN (see Section \ref{sec:interloper} of the Appendix for more detail), whose variable nature makes it difficult to draw conclusions on the net overall flux from AT2019cuk. However, the Swift XRT and UVOT data, along with the ground-based optical data, also do not show any significant periodic variability on timescales of roughly 30 days. The lack of periodic behavior across the electromagnetic spectrum suggests that the initial behavior found in ZTF was due to stochastic red noise variability commonly observed in AGN. Despite the lack of periodic behavior, the large increase in the optical and infrared flux around 2018 and the corresponding change in the optical Balmer lines still make this a very compelling changing-look AGN.

\subsection{Comparison to Other Nuclear Transients}

\subsubsection{Quasi-Periodic Eruptions (QPEs)}

Flares on the order of hours-days are also seen in the recently discovered QPEs \citep{Miniutti2019,Giustini2020,Arcodia2021,Chakraborty2021}. However, there are numerous differences between the hard X-ray flares in AT2019cuk and QPEs, including that QPEs are primarily soft X-ray emitting sources, show much larger amplitude flares, and return to a stable low X-ray luminosity phase in between flares. Thus, the flares seen in AT2019cuk do not appear QPE-like. 

\subsubsection{Changing-Look AGN}

The optical spectrum of AT2019cuk has shown clear changes since the SDSS spectrum taken in 2005 with a significant change to the broad H$\alpha$ line and the appearance of a broad component of the H$\beta$ complex \citep{Jiang2022}, which is the defining signature of a changing-look AGN. Similar to AT2019cuk, an accompanying rise in optical flux has been seen in numerous changing-look AGN with newly emerging broad lines \citep[e.g.][]{Shappee2014,MacLeod2016,Trakhtenbrot2019}, which is commonly attributed to an increase in the mass accretion rate, although the mechanism driving this change is still debated \citep[see][and references therein]{Ricci2022}. In the X-ray band, many changing-look AGN exhibit spectral changes, but most are not outside of the typical spectral variations seen in standard AGN \citep[e.g.][]{Oknyansky2019,Parker2019,Wang2020,Guolo2021}. Unfortunately, there are no X-ray observations that pre-date the optical outburst of AT2019cuk, except for ROSAT survey upper limits, so we cannot comment on the long-term X-ray variability. However, while AT2019cuk often looks like a relatively standard AGN in its non-flaring state, it shows the first evidence for day-long, hard X-ray flares in a changing-look AGN. The high-cadence NICER observations of AT2019cuk were crucial to the discovery of these hard X-ray flares and motivate more high-cadence X-ray monitoring of changing-look AGN to determine whether such X-ray behavior is common in these systems. 
 
\section{Summary} \label{sec:concl}

During a high-cadence monitoring program of the binary SMBH candidate and changing-look AGN AT2019cuk with NICER, we discovered hard X-ray flares unlike typical AGN flares. The flares appear most prominently in the 2-4~keV band, are roughly day-long, and are repetitive in nature, although they do not show signs of periodicity. We performed spectral modeling of all of the NICER data on a GTI-by-GTI basis, using both a blackbody and ionized absorber for the soft excess found in the XMM-Newton data. The flares have anomalously hard X-ray spectra with significantly lower photon indices than are typically seen in AGN and the non-flaring data. We present three possible models, namely corona/jet variability driven by increased magnetic activity, variable obscuration, and self-lensing in a binary SMBH, for these hard X-ray flares. We favor the corona variability model, given that the obscuration model requires significant fine-tuning of the timescales involved and the self-lensing model is difficult to reconcile with aperiodic flares. Finally, the multi-wavelength monitoring presented here shows no signs for 30~day periodicity, as was first reported in \cite{Jiang2022}.

\vspace{0.5cm}

We thank the referee for feedback that helped improve this work. DRP was partially funded by the NASA grant 80NSSCK0962. DJD received funding from the European Union's Horizon 2020 research and innovation programme under Marie Sklodowska-Curie grant agreement No. 101029157, and from the Danish Independent Research Fund through Sapere Aude Starting Grant No. 121587. ECF is supported by NASA under award number 80GSFC21M0002. MI, GSHP acknowledge the support from the National Research Foundation of Korea grants, No. 2020R1A2C3011091 \& No. 2021M3F7A1084525, and the Korea Astronomy and Space Science Institute grant under the R\&D program (Project No. 2022-1-860-03) supervised by the Ministry of Science and ICT. 

We thank the staff of CBNUO and LOAO, Hyori Jeon, Haeun Kim, Jeongmuk Kim, Chaerin Kim, Jae-Hyuk Yoon, and In-Kyung Paek, for their help with the optical observations. This work made use of the data obtained at CBNUO and LOAO. LOAO is operated by the Korea Astronomy and Space Science Institute.


\facilities{NICER, XMM-Newton, NuSTAR, Swift, ZTF, LOAO, MAO, CBNUO}

\software{XSPEC (v12.12.0; \citealt{Arnaud1996})
HEASoft (v6.30.1; \citealt{HEASARC2014})
XMM-SAS (v20.0.0; \citealt{Gabriel2004})
HOTPANTS \citep{Becker2015}}

\newpage

\appendix

\section{Verifying the Hard X-ray Flares} \label{sec:verify}

\subsection{NICER Data Quality Checks} \label{sec:verify_nicer}

Given the decreasing effective area of NICER at high energies and uncertainties in background modeling for NICER data, the hard X-ray flares seen in AT2019cuk require extreme scrutiny to ensure that they are neither instrumental nor a result of background contamination. In addition, periods of high background flaring may become more prevalent as we approach solar maximum and experience periods of increased solar activity, whereas NICER calibration data were primarily acquired early in the mission near solar minimum. Thus, we take numerous precautions in addition to the relatively restrictive data filtering presented in Section \ref{subsec:nicer} to verify that these flares are not instrumental. 

\begin{figure}[t]
    \centering
    \includegraphics[width=\textwidth]{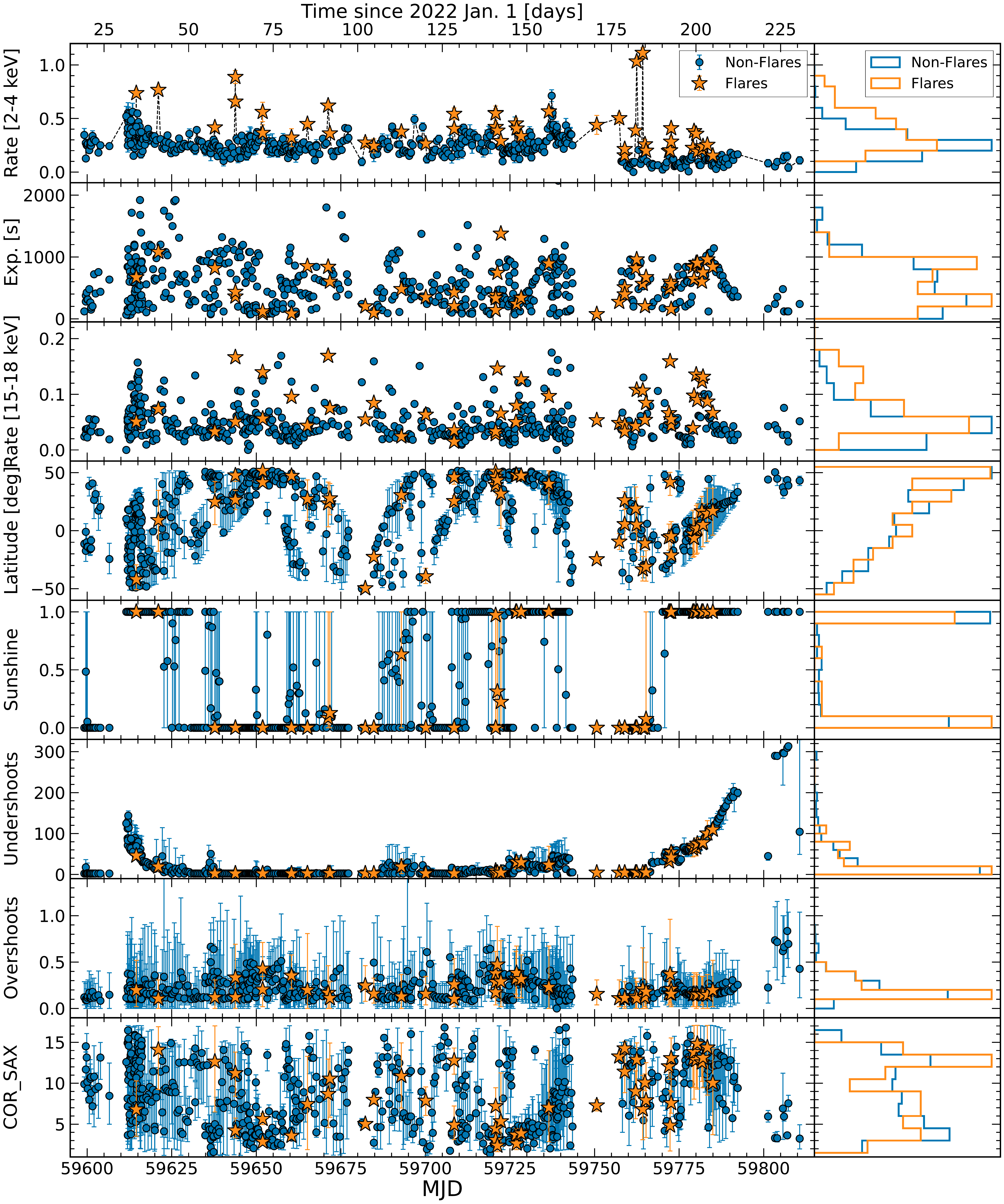}
    \caption{NICER instrument parameters for the flares and non-flares. The rows correspond to: (1) Hard count rate (2-4 keV), (2) Exposure time for the GTI, (3) Raw high energy count rate (15-18 keV), (4) Satellite latitude, (5) NICER sunshine factor, (6) Undershoots, (7) Overshoots, (8) Cut-off rigidity. In all 8 panels, the blue circles correspond to non-flaring data points and orange stars correspond to flares, as defined in Section \ref{subsec:flares}. In the lower five panels, corresponding to NICER instrumental parameters, the error bars indicate the minimum and maximum values of each parameter reached during a given GTI. The right-most panels are normalized distributions for the non-flares in blue and the flares in orange for each of the 8 parameters previously described. No clear difference in seen between flares and non-flares, indicating that none of these NICER instrument parameters are responsible for the flaring behavior.}
    \label{fig:nicer_checks}
\end{figure}

We first ensure that these flares are not a result of the ISS being too close to either the polar horn or the South Atlantic Anomaly (SAA) regions of Earth's magnetosphere, both of which are known to produce significant background counts indistinguishable from cosmic X-rays. The flares are distributed throughout the orbital path of the satellite and are not preferentially located near either of these problematic regions. Next, we checked numerous other quantities associated with each GTI, including the exposure time, background rate, high-energy (15-18 keV) raw count rate, sunshine factor, and cutoff rigidity factor (COR), to compare the distribution of these quantities between flares and non-flares. We find that there is no significant difference between the distributions of any of these factors in flaring versus non-flaring GTIs. In addition, we check the undershoots and overshoots for each GTI and find that the flares and non-flares follow the same distributions and are reasonably distributed below the normal filtering levels. Additionally, the overshoot rates also fall below the COR-dependent value from the \texttt{overonly\_expr} term in standard \texttt{nicerl2} filtering. In Figure \ref{fig:nicer_checks}, we show the evolution of many of these parameters used in checking the validity of the hard flares, with a distribution of the non-flares and flares for each parameter to the right, each of which show consistency between the two populations.

Finally, another concern is that many of the flares occurred in only single, isolated GTIs. Although there is nothing in the data quality checks detailed above that hints at issues with these data points, we also checked for consistency among isolated flares (one GTI surrounded by non-flaring GTIs on either side) and non-isolated flares ($>2$ GTIs in a row marked as a flare). We find that there are no differences between the isolated and non-isolated flares, either in the data quality checks presented in Figure \ref{fig:nicer_checks} or in the spectral fitting results. This suggests that the two are not independent distributions and the isolated flares can be treated as just as valid as the non-isolated flares. Likewise, we note that the number of isolated flares was greatest in the first 100 days or so of the NICER monitoring, in which the sampling rate was lower.

\subsection{Simultaneous Swift Observations} \label{sec:verify_swift}

\begin{figure}[t]
    \centering
    \includegraphics[width=\textwidth]{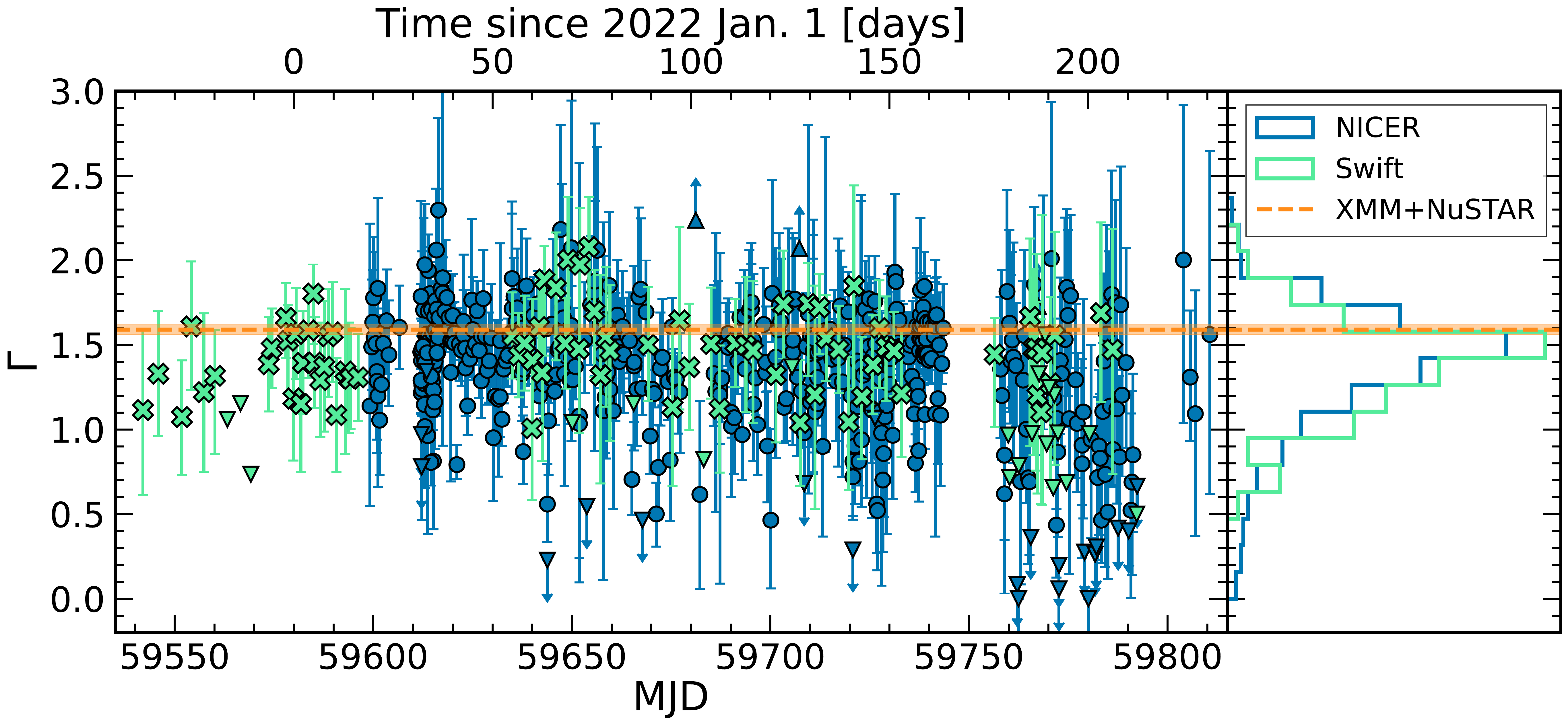}
    \caption{\textit{Left:} Distribution of photon indices as a function of time for both NICER (blue) and Swift XRT (green) observations. Both sets of data are fit with an absorbed blackbody plus power law model, but in the Swift fitting, the column density of the absorber and the blackbody temperature are both fixed at the value of the XMM-Newton/NuSTAR fits due to limited spectral constraints. Any upward/downward facing triangles indicate lower/upper limits, respectively, where the upper limit of $\Gamma$ was set at 3 and the lower limit was set at 0. \textit{Right:} Histogram of photon indices across all observations with both NICER (blue) and Swift (green). The distributions are similar, with both extending down to quite low photon indices compared to standard AGN. In both panels, the orange dashed line is the XMM-Newton/NuSTAR value for the same model, with the shaded area showing the uncertainty on that value.}
    \label{fig:gamma_swift+nicer}
\end{figure}

Given the extensive Swift observations of AT2019cuk, we also check our NICER spectral modeling results by comparing to Swift spectral modeling. Swift has imaging capabilities that NICER lacks, which is beneficial because it allows us to ensure that the results are not driven by a nearby AGN (which is further discussed in Section \ref{sec:interloper}) nor the uncertainty in the empirical NICER background. Swift spectra and background spectra were created on a per-ObsID basis, using the online Swift XRT products generator \citep{Evans2007,Evans2009}. The spectra were then binned using the optimal binning procedure \citep{Kaastra2016} and to a minimum of 1 count per bin. We fit each spectrum with $C$ statistic minimization, given the limited spectral quality and lower count rate of Swift relative to NICER. We fit the Swift spectra in the 0.3-10~keV range to the model \texttt{tbabs*ztbabs*(zpo+zbb)}, with the blackbody temperature fixed to the XMM-Newton/NuSTAR value of $kT_\mathrm{bb} = 0.081$~keV and the column density of the neutral absorber fixed to the XMM-Newton/NuSTAR value of $N_H = 6.9 \times 10^{20}$~cm$^{-2}$. We fixed these parameters to eliminate any degeneracies with the photon index due to the limited spectral quality of Swift XRT data. The resulting photon indices as a function of time are shown in Figure \ref{fig:gamma_swift+nicer}, with a comparison to all NICER fits as described in Section \ref{subsec:spec}.

We find that the Swift and NICER data have very similar distribution of photon indices, albeit with large error bars on the Swift data. Some of the Swift data points have relatively low photon indices with $\Gamma \lesssim 1$, which is similar to what we find in the flaring NICER data with the same model. Thus, we find similar evidence in the Swift data for anomalously low photon indices and hard X-ray spectra, indicating that the flares seen in NICER are not a result of the contaminating source nor a result of poorly understood NICER background.

There is some evidence for a lower photon index preferentially found in the Swift data. This has two possible explanations--(1) NICER data are contaminated by the nearby AGN that has a softer spectrum with $\Gamma \approx 1.7$ (see Section \ref{sec:interloper}) or (2) Swift spectra are a combination of both flaring and non-flaring time periods, given the longer baseline in each data point due to the per-ObsID sampling. The first scenario is disfavored because the XMM-Newton/NuSTAR spectra, which also have the ability to clearly resolve the two sources, show a similar photon index to the average NICER value (when in a non-flaring state). Thus, we suspect that the lower photon indices seen in the Swift modeling are a result of coarser binning than we use with NICER, which could lead to some Swift data points including both flaring and non-flaring time periods. We require this rather coarse, ObsID-based binning rather than per-GTI binning with Swift in order to have sufficient signal to perform spectral fitting. However, when we bin the Swift light curve on a per-GTI basis (similar to NICER binning), we find that there are similar looking flares in the Swift light curve (see the bottom two panels of Figure \ref{fig:19cuk_swift+nicer}).

\subsection{Comparison to Simulated NICER Spectra} \label{subsec:verify_sims}

\begin{figure*}[t]
    \centering
    \includegraphics[width=\textwidth]{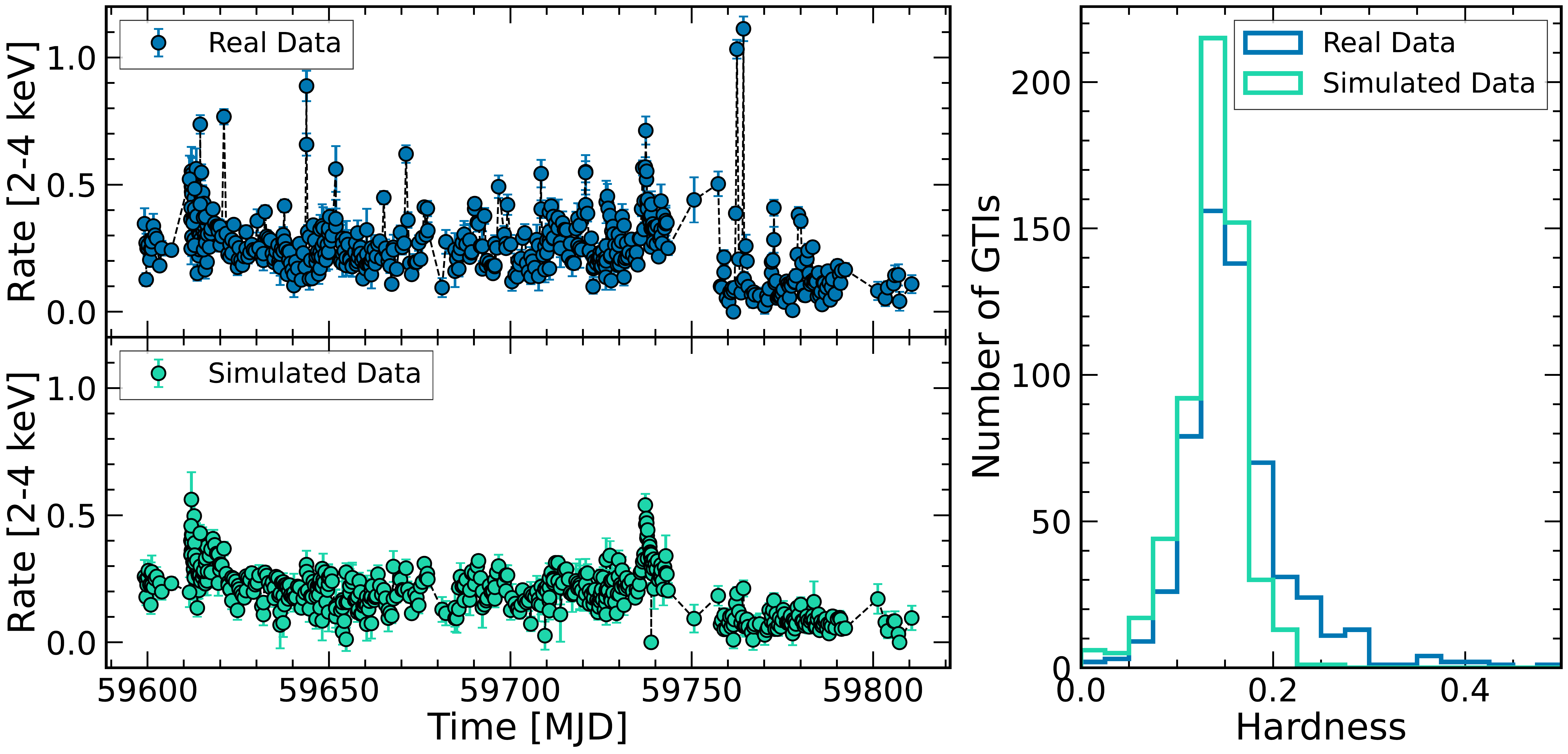}
    \caption{\textit{Left:} GTI-based hard X-ray light curves (2-4~keV) for both the real NICER observations (top) and the simulated NICER observations (bottom) with the same distribution exposure times as the real observations. The simulated data are based on the joint XMM-Newton/NuSTAR fit with the partial covering warm absorber. The real data show clear, strong flares that are not present in the simulated data, indicating that the flares are not due to noise induced from counting statistics. \textit{Right:} Distribution of hardness ratios for both the real (dark blue line) and simulated (light green line) data. There is a clear skew in the real data toward harder spectra.}
    \label{fig:sim_nicer}
\end{figure*}

The NICER observations are also rather short and therefore contain a relatively low number of counts compared to the longer XMM-Newton and NuSTAR observations. Hence, to check whether the flares could arise from noise induced by counting statistics, we simulated a similar distribution of NICER observations from the best fit to the XMM-Newton/NuSTAR with the partial covering warm absorber. As each of the NICER GTIs has a different flux, exposure time, and background spectrum, we utilize the properties of each GTI to produce the simulated data, including rescaling the best fit XMM-Newton/NuSTAR spectrum to the flux of the NICER GTI. The simulated observations do not reveal any strong hard X-ray flares in the 2-4~keV band, as are seen in the real data, which can be seen qualitatively in the left panel of Figure \ref{fig:sim_nicer} where we plot both the real and simulated hard X-ray light curves. In addition, the rightmost panel of Figure \ref{fig:sim_nicer} shows that the hardness ratio seen in the real data is significantly harder than the simulated hardness ratio. Thus, the flares are not a result of noise due to counting statistics and cannot be attributed to standard AGN variability.

\section{Impact of a Nearby AGN on NICER Data} \label{sec:interloper}

\begin{figure}[t]
    \centering
    \includegraphics[width=18cm]{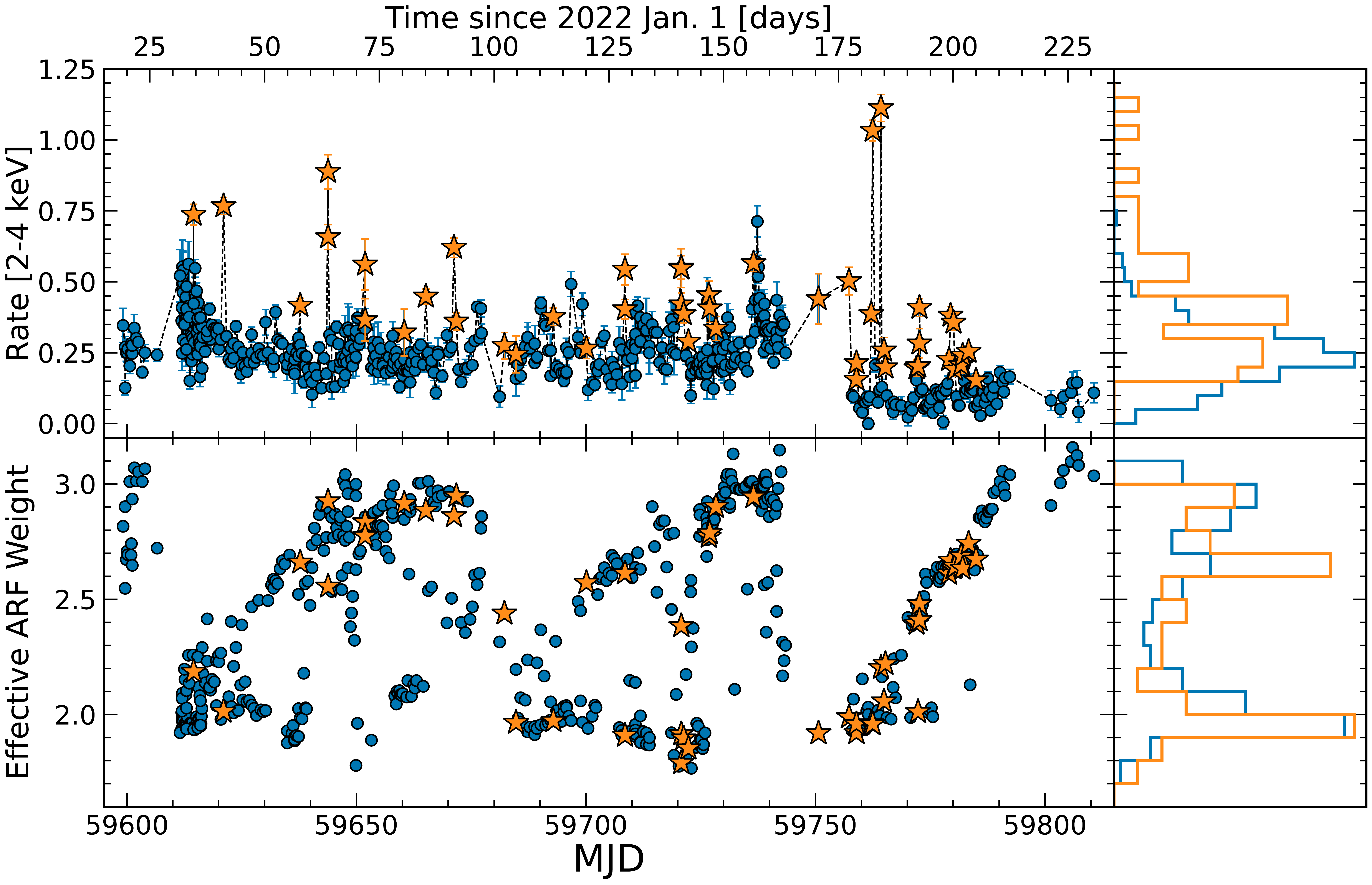}
    \caption{\textit{Top panel:} NICER 2-4 keV light curves for the AT2019cuk field of view. \textit{Bottom panel:} Effective ARF weight, defined as the mean of the weighting for each detector at the location of AT2019cuk divided by the mean of the weighting for each detector at the location of the interloper, both of which are computed using \texttt{nicerarf}. In both panels the blue circles represent non-flaring GTIs and the orange stars represent flaring GTIs, defined by the hardness ratio cut detailed in Section \ref{subsec:flares}. The right side shows histograms for both the 2-4 keV rate and the effective ARF weight for the two different populations (flaring vs. non-flaring) in their respective colors. As the flares and non-flares have similar effective ARF weights, it is evident that the flares are not a result of pointing differences or jitter.}
    \label{fig:arfweight}
\end{figure}

Upon investigating AT2019cuk with X-ray imaging telescopes including XMM-Newton, NuSTAR, and Swift, we noticed a nearby strong X-ray source located approximately 3.1' away from AT2019cuk. This interloping source is a known AGN, located at ($\alpha$,$\delta$) = (14:30:08.65, +23:06:21.62) \citep{Veron-Cetty2010}. At 3.1' away from AT2019cuk, this interloper is just within the field of view of NICER, but at a location such that the detector response has dropped by a factor of 2-3 in sensitivity, with the exact response depending on the exact pointing and which detectors are used. We compute the relative response of NICER's X-ray Timing Instrument at the location of this interloper by using \texttt{nicerarf} with the RA and Dec set to the position of the interloper. We can then define the ``effective ARF weight" as the mean of the weighting for each detector at the location of AT2019cuk divided by the mean of the weighting for each detector at the location of the interloper. A high effective ARF weighting means that AT2019cuk dominates the flux in the NICER pointings. In Figure \ref{fig:arfweight}, we show this effective ARF weighting over the entire observing period, together with the 2-4 keV light curve for easy comparison with the flares. From this it is clear that the flares are not simply an effect of pointing oscillations, as they appear in both times of high and low effective ARF weighting, with a similar distribution to the non-flaring points.

We also extracted a NuSTAR spectrum for both AT2019cuk and the interloper from the DDT observation in January 2022 to compare the spectral shapes between AT2019cuk and the interloper. Both spectra can be fit well with an absorbed power law. We leave the photon indices free between the two sources, and the best fit gives $\Gamma = 1.74_{-0.07}^{+0.08}$ for AT2019cuk and $\Gamma = 1.77_{-0.14}^{+0.17}$ for the interloper. Thus, as the spectral shape of the interloper is extremely similar to that of AT2019cuk and we see significant changes in the spectral shape during the hard X-ray flares, we do not expect the flares to be a result of the interloper.

\begin{figure}[t]
    \centering
    \includegraphics[width=\textwidth]{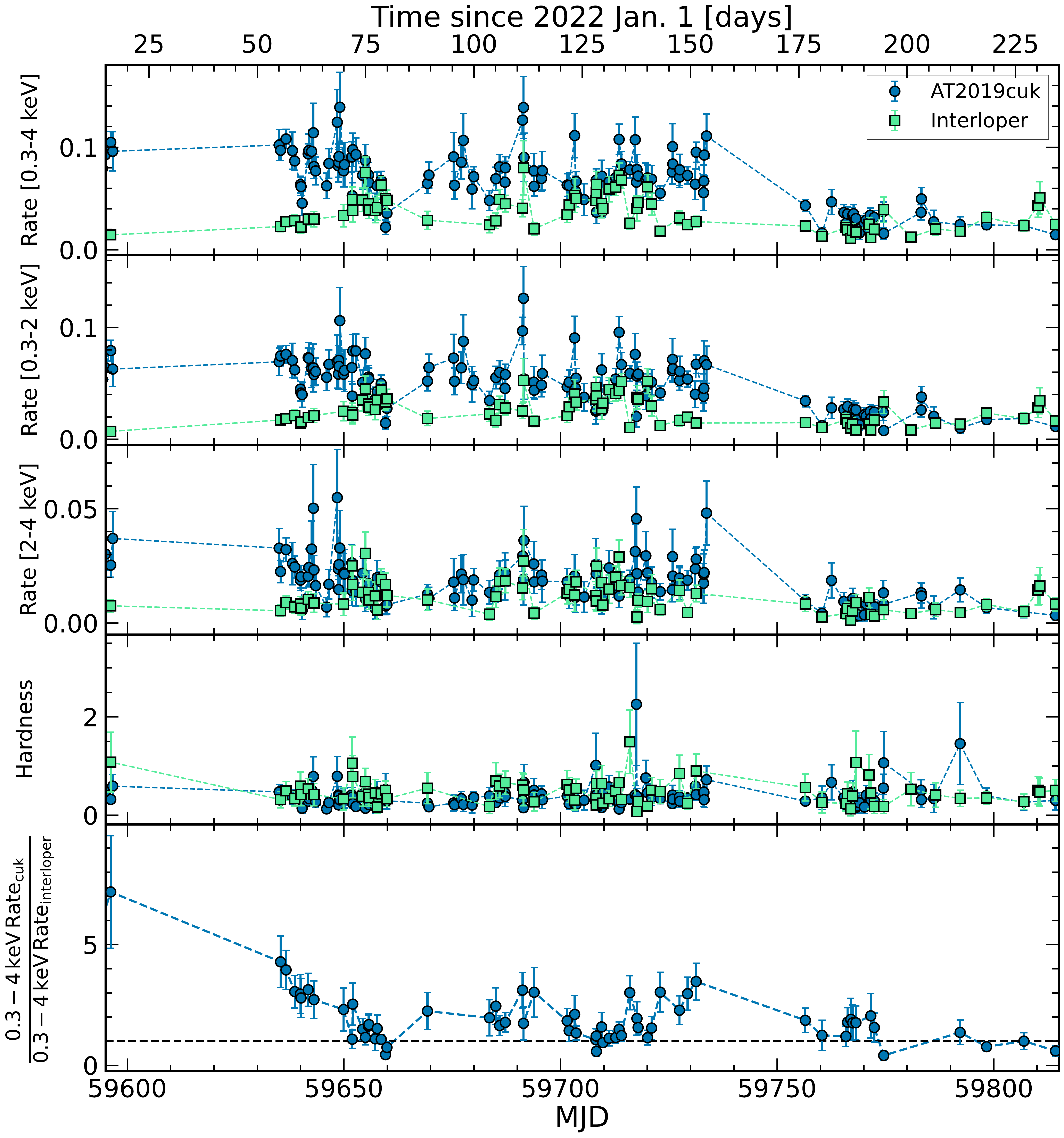}
    \caption{\textit{Top 3 panels:} Swift XRT light curves in three different energy bands (from top to bottom: 0.3-4 keV, 0.3-2 keV, and 2-4 keV). Each light curve is grouped by GTI, is background-subtracted, and shown in counts per second. The blue circles and green squares correspond to AT2019cuk and the interloper, respectively. \textit{Fourth panel:} Swift hardness ratio, defined as the count rate in 2-4 keV divided by the count rate in 0.3-2 keV. The two sources have relatively comparable hardness ratios, meaning that the hard X-ray flares seen in NICER are not likely due to picking up more flux from the interloper. \textit{Bottom panel:} Ratio of the 0.3-4 keV count rates of AT2019cuk and the interloper. The black dashed line shows an equal 0.3-4 keV rate between the two sources, and any data points above the line indicate that AT2019cuk was brighter than the interloper. In combination with the difference in the response at the location of the interloper (see Figure \ref{fig:arfweight}), the bottom panel shows decisively that the interloper contributes at worst roughly half of the observed NICER rate for AT2019cuk.}
    \label{fig:interloper_swift}
\end{figure}

Beyond the impact of the interloper on the NICER flares, we also need to assess the impact of the interloper on the overall variability of AT2019cuk. This is especially important for assessing any long-term periodicity of the light curve that could be suggestive of a binary SMBH. Given their ability to distinguish between the two sources, we used both NuSTAR and Swift to investigate the influence of the interloper relative to AT2019cuk. In the initial NuSTAR observation in January 2022, the interloper was a factor of three lower in flux than AT2019cuk. In Figure \ref{fig:interloper_swift}, we show the Swift light curve for these two sources during the NICER campaign, including a ratio of their count rates in the 0.3-4 keV band in the bottom panel. For the majority of the observing campaign, the interloper is brighter than AT2019cuk in the 0.3-4 keV band, although there are some time periods in which the interloper is brighter by a factor of no more than two. However, combined with the drop in the effective response at the location of the interloper of at least two, we find that the interloper never contributes more than 50\% of the flux in the NICER band and usually contributes rather negligibly. Given the variability in the amount of flux from the interloper and the large gaps in Swift monitoring of this source during the NICER campaign, it is difficult to assess exactly how much impact this has on the overall variability of AT2019cuk with NICER.

\begin{figure}[t]
    \centering
    \includegraphics[width=\textwidth]{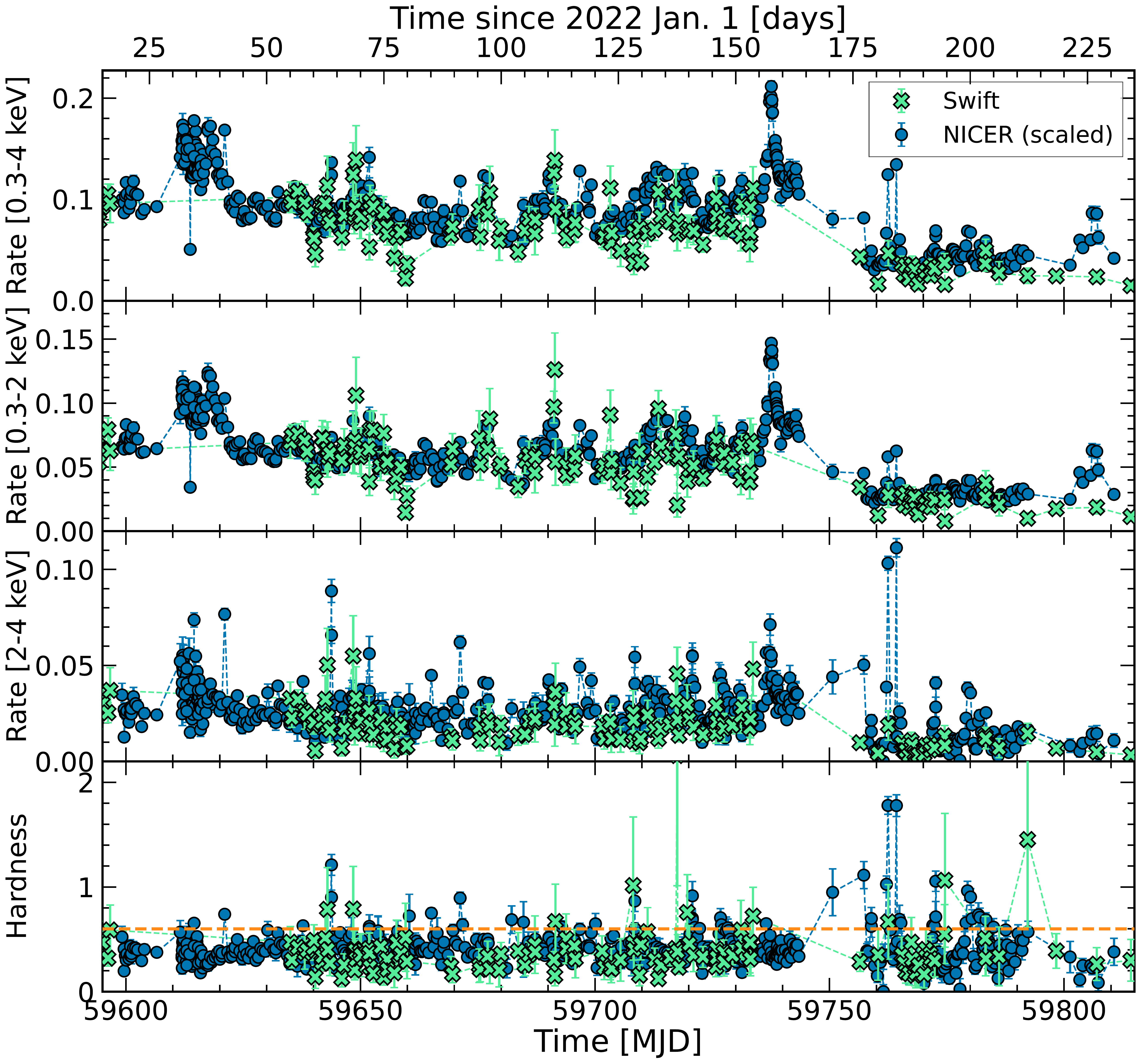}
    \caption{\textit{Top 3 panels:} NICER (blue circles) and Swift XRT (green X's) light curves in three different energy bands (from top to bottom: 0.3-4 keV, 0.3-2 keV, and 2-4 keV) for AT2019cuk. Both the NICER and Swift data are binned by GTI. Each light curve is background-subtracted and show in counts per second. \textit{Bottom panel:} NICER and Swift hardness ratios, defined as the count rate in 2-4 keV divided by the count rate in 0.3-2 keV. The orange dashed line shows the corresponding hardness ratio cut-off for the flares, defined in Section \ref{subsec:flares}. In each panel, the NICER data have been rescaled by a constant factor to account for the effective area difference between the two instruments. There is generally good agreement between the Swift and NICER data after this rescaling, which indicates that AT2019cuk is contributing most of the flux that NICER sees, rather than the interloper. Similarly, there are some relatively hard Swift data points coincident with some of the NICER flares. }
    \label{fig:19cuk_swift+nicer}
\end{figure}

In addition, we can utilize the Swift monitoring of AT2019cuk to assess the effect of the interloper on the NICER data. In Figure \ref{fig:19cuk_swift+nicer}, we show both the NICER and the Swift light curves of AT2019cuk, where in each of the panels the NICER light curves have been scaled by a constant factor to account for the difference in the effective area between NICER and Swift. These factors were determined using the HEASARC WebPIMMS tool\footnote{\url{https://heasarc.gsfc.nasa.gov/cgi-bin/Tools/w3pimms/w3pimms.pl}} and assuming a $\Gamma = 1.7$ power law for the conversion between the different count rates. In the 0.3-4 keV, 0.3-2 keV, and 2-4 keV bands respectively, these renormalization factors are 20, 25, and 10, where these factors represent the amount by which we divided the NICER data to accurately compare to the Swift data. From Figure \ref{fig:19cuk_swift+nicer}, we can see that the NICER variability matches the Swift variability of AT2019cuk very well, including picking up key features of the light curve like the drop in flux around MJD 59750. Thus, this indicates that most of the variability in the NICER data is coming from AT2019cuk rather than the interloper.

\bibliographystyle{aasjournal}
\bibliography{refs}

\end{document}